\documentclass[11pt]{article}
\usepackage{amssymb}
\usepackage{graphics}
\usepackage{epsfig}
\usepackage{a4wide}
\usepackage{slashed,cite}
\usepackage{color}

\begin{document}

\newcommand{\nn}{\nonumber}
\newcommand{\eezrr}{$e^+e^- \to Z \gamma\gamma~$}

\title{Diphoton plus $Z$ production at the ILC at ${\cal O}(\alpha^4)$ }
\author{ Zhang Yu, Guo Lei, Ma Wen-Gan, Zhang Ren-You, Chen Chong, and Li Xiao-Zhou   \\
{\small  Department of Modern Physics, University of Science and Technology of China (USTC),}  \\
{\small   Hefei, Anhui 230026, P.R.China}}

\date{}
\maketitle \vskip 15mm
\begin{abstract}
Precision measurement for the production of a $Z$-boson in
association with two photons is important for investigating the Higgs
boson and exploring new physics at the International Linear Collider. It could be used to study the $ZZ\gamma\gamma$ anomalous
quartic gauge coupling. In this work we report on our calculation of
the full ${\cal O} (\alpha^4)$ contributions to the $e^+e^- \to Z
\gamma\gamma~$ process in the standard model, and we analyze the
electroweak (EW) quantum effect on the total cross section. We
investigate the dependence of the $Z\gamma\gamma$ production rate on
the event selection scheme and provide distributions for some
important kinematic observables. We find that the next-to-leading
order (NLO) EW corrections can enhance the total cross section
quantitatively from $2.32\%$ to $9.61\%$ when the colliding energy goes
up from $250~GeV$ to $1~TeV$, and the NLO EW corrections show
obviously a nontrivial phase space dependence. We conclude that in
studying the signal process $e^+e^- \to ZH \to Z \gamma\gamma~$, the
background process $e^+e^- \to Z \gamma\gamma~$ can be suppressed
significantly if we take appropriate kinematic cuts on the final
products.
\end{abstract}

\vskip 15mm {\large\bf PACS: 13.66.-a, 14.70.Hp, 14.70.Bh }

\vfill \eject \baselineskip=0.32in

\renewcommand{\theequation}{\arabic{section}.\arabic{equation}}
\renewcommand{\thesection}{\Roman{section}.}
\newcommand{\nb}{\nonumber}

%slash:
\newcommand{\Dir}{\kern -6.4pt\Big{/}}%su lettere italiane minuscole
\newcommand{\Dirin}{\kern -10.4pt\Big{/}\kern 4.4pt}
\newcommand{\DDir}{\kern -7.6pt\Big{/}}%su lettere italiane maiuscole
\newcommand{\DGir}{\kern -6.0pt\Big{/}}%su lettere greche

\makeatletter      % '@' is now a normal "letter" for TeX
\@addtoreset{equation}{section}
\makeatother       % '@' is restored as a "non-letter" character for TeX

\par
\section{Introduction}
\par
Probing the mechanism of electroweak symmetry breaking (EWSB) is one
of the most important tasks in particle physics. In the standard
model (SM), symmetry breaking is achieved by introducing the
Higgs mechanism, which gives masses to the elementary particles and
implies the existence of an SM Higgs boson. Therefore, to uncover
the origin of EWSB and to determine whether the SM Higgs boson
really exists is one of the highlights of the Large Hadron Collider
(LHC) physics program \cite{LHC}. In July 2013, both the ATLAS and CMS
collaborations at the LHC reported that they had observed a new
neutral boson with mass of around $126~GeV$ \cite{CMS,ATLAS}, and
this particle is tentatively identified as a Higgs boson. The more precise measurements on its properties are still
going on at the LHC, but in light of the current data, its
properties are very well compatible with the SM Higgs boson.
However, it has been understood for a long time that there are
intrinsic limitations from the ability of hadron colliders in
precision measurement. The International Linear Collider (ILC) is an
ideal machine to address this problem \cite{ILC}. One of the major
aspects of the physics program of the ILC is to make detailed
precision measurements of the nature of the Higgs boson discovered
at the LHC \cite{ILC,ILC-higgs}.

\par
For any observed new particle, the determination of its fundamental
properties will be a primary goal. The measurement of the branching
fraction of the Higgs boson decaying into two photons,
$Br(H\to\gamma\gamma)$, turns out to be an absolutely necessary
ingredient in extracting the total width \cite{hrr-john}. Besides,
this measurement may possibly provide hints for new physics if the
deviation from the SM prediction is larger than the measurement
accuracy. At the ILC the Higgs boson is predominantly produced by
the Higgs-strahlung process $e^+e^-\to ZH$. The most serious and
irreducible background for Higgs search via the $H \to \gamma \gamma$
decay channel arises from the \eezrr process, which it is hard to 
get rid of and needs to be explored in depth \cite{hrr}.

\par
The precision measurement of the quartic gauge boson coupling (QGC)
can provide a connection to the mechanism of electroweak symmetry
breaking. The anomalous QGC, such as $ZZ\gamma\gamma$, vanishes in
the SM at the tree level and might provide a clean signal of new
physics, since any deviation from the SM prediction might be
connected to the residual effect of electroweak symmetry breaking.
The effect of $ZZ\gamma\gamma$ coupling has been theoretically
investigated at the LEP and the ILC \cite{Belanger:1992,
Belanger:2000, Stirling:2000, Stirling:1999, Montagna:2001,
Villa:2005}. The measurement of the \eezrr process at LEP2 by L3
Collaboration \cite{LEP:2002} shows that the anomalous
$ZZ\gamma\gamma$ coupling leads to a negligible effect at LEP energy,
while it might be detectable at the ILC with higher colliding
energy. Since this effect can be small and subtle, theoretical
predictions for the cross section with high precision is mandatory.

\par
At the ILC, the accuracy of the cross section measurement for the
triple gauge bosons production process could reach per mille level.
It is necessary to presume upon an accurate theoretical calculations
to match the experimental accuracy. Thus a good theoretical
predictions beyond leading order (LO) are indispensable. In the last
few years, a lot of work contributed to the phenomenological
studies in the SM up to the QCD next-to-leading order (NLO) on
triple gauge boson production processes at hadron colliders
\cite{Lazopoulos:2007ix, Hankele:2007sb, Campanario:2008yg,
Binoth:2008kt, Bozzi:2009ig, Bozzi:2010sj, Baur:2010zf,
Bozzi:2011ww,Bozzi:2011zrr}. Most recently, the calculation of the
NLO electroweak (EW) correction to the $W^+W^-Z $ production at the
LHC was present in Ref.\cite{wwz-ew}. The NLO EW calculations to the
$W^+W^-Z $ and $ZZZ$ productions at the ILC were provided in
Refs.\cite{sjj,sw,sh}, while a prediction for the $Z$ production
associated with two photons at the ILC in the NLO EW precision is
still missing.

\par
In this paper, we investigate the complete NLO EW corrections to the
\eezrr process at the ILC in the SM. The rest of the paper is
organized as follows: In the following section we present the LO and
NLO EW analytical calculations for the \eezrr process. The
numerical results and discussions are given in Section III. Finally,
we will give a short summary.

\vskip 5mm
\section{Analytical calculations}
\par
The LO and NLO EW calculations for the \eezrr process in the SM are
presented in this section by using the 't Hooft-Feynman gauge. We
apply FeynArts-3.7 package \cite{feynarts} to automatically generate
the Feynman diagrams and the FormCalc-7.2 program \cite{formcalc} to
algebraically simplify the corresponding amplitudes. In our
calculations we neglect the contributions of the Feynman diagrams
which involve $H$-$e$-$\bar{e}$, $G^0$-$e$-$\bar{e}$,
$G^+$-$e$-$\bar{\nu}_{e}$ or $G^-$-$\nu_{e}$-$\bar{e}$ vertices,
because the Yukawa coupling strength of Higgs/Goldstone to fermion
pair is proportional to the fermion mass.

\par
\begin{figure*}
\begin{center}
\includegraphics [scale = 1.0]{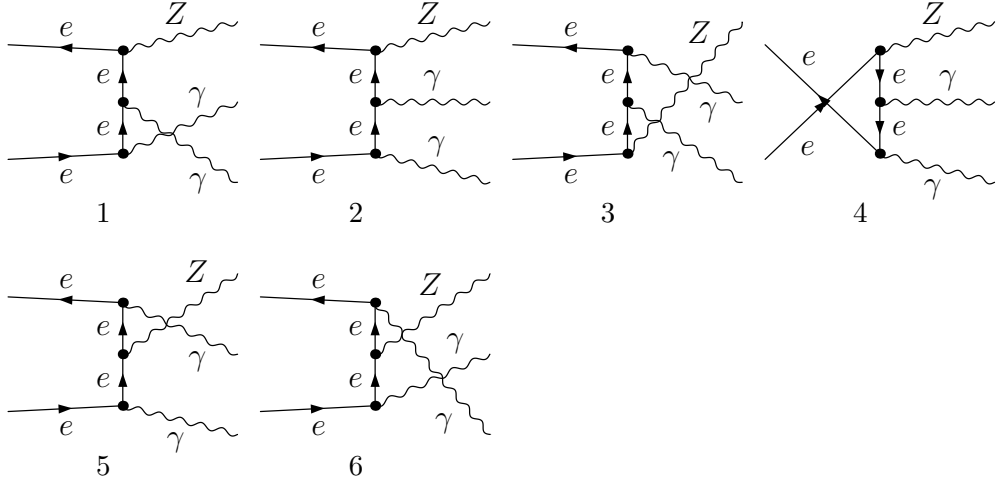}
\caption{\label{fig1} The tree-level Feynman diagrams for the \eezrr
process. The graphs with the exchange of the final two photons are not
drawn. }
\end{center}
\end{figure*}

We denote the process 
\begin{eqnarray}\label{eq-lo}
e^+(p_1)+e^-(p_2)\to Z(p_3)+\gamma(p_4)+\gamma(p_5),
\end{eqnarray}
where $p_i(i=1,5)$ represent the four-momenta of the initial and
final particles. There are six generic tree-level Feynman diagrams
for the \eezrr process, and some of them are depicted in
Fig.\ref{fig1}. The LO cross section for the \eezrr process can be
obtained as
\begin{eqnarray}\label{eq-cslo}
\sigma_{LO}=\frac{1}{2!}\frac{(2\pi)^4}{2s}\int d\Phi_{3}
\overline{\sum_{spin}} |{\cal M}_{LO}|^2,
\end{eqnarray}
where ${\cal M}_{LO}$ is the LO amplitude, the factor $\frac{1}{2!}$
comes from the two identical final photons and the bar over
summation recalls averaging over the initial spins. The phase space
element of the final three particles is defined as
\begin{eqnarray}
{d\Phi_3}=\delta^{(4)} \left( p_1+p_2-\sum_{i=3}^5 p_i \right)
\prod_{i=3}^5 \frac{d^3 \vec{p}_i} {(2 \pi)^3 2 E_i}.
\end{eqnarray}
%%%%%%%%%%%%%%%%%%%%%%%%%%%%%%

\par
\begin{figure*}
\begin{center}
\includegraphics [ scale = 1.0]{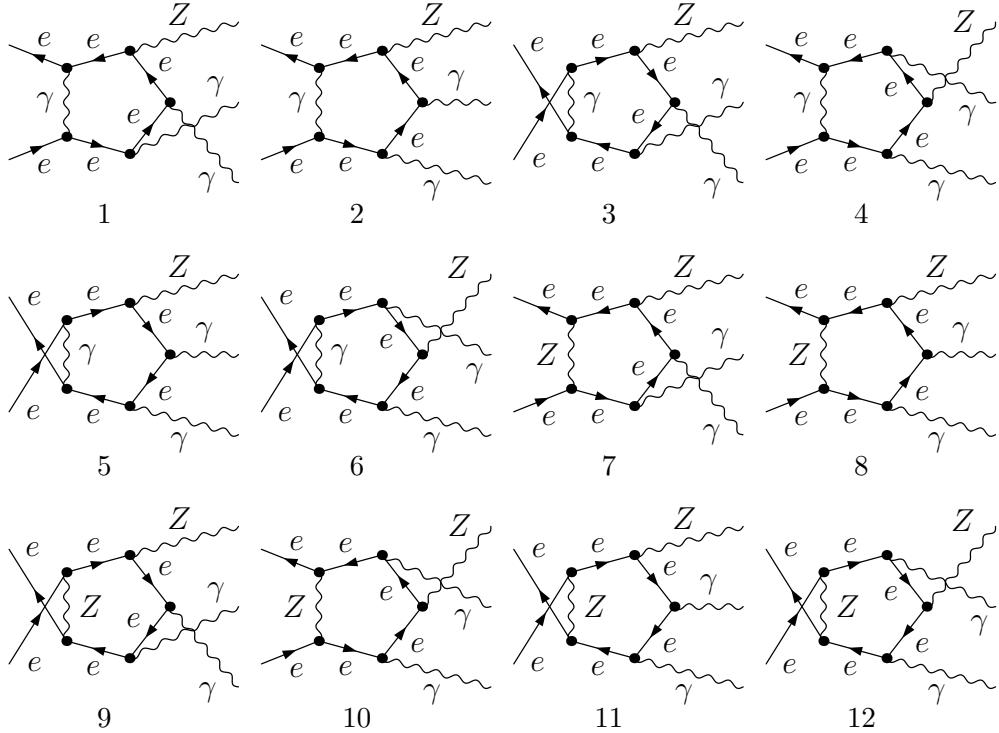}
\caption{\label{fig-pen} The pentagon diagrams for the \eezrr
process which are calculated by using the codes with quadruple
precision. The diagrams with exchanging the final two photons are
not drawn. }
\end{center}
\end{figure*}

The virtual EW correction to the \eezrr process at ${\cal
O}(\alpha^4)$ involves $1003$ diagrams, including $36$ self-energy
diagrams, $472$ triangles, $418$ boxes, $47$ pentagons and $30$
counterterm graphs. The most complicated topology involved in the EW
one-loop contribution contain $5$-point integrals up to rank $4$,
which are deduced by using the reduction method in
Ref.\cite{denner}. The numerical calculations of $n$-point $(n \le
4)$ tensor integrals are implemented by using the Passarino-Veltman
reduction algorithm \cite{PV}. We adopt mainly the LoopTools-2.8
package \cite{formcalc} for the numerical calculations of the scaler
and tensor integrals. In order to avoid instability in the numerical
calculations of the 5-point tensor integrals of rank $4$, we developed
the program coded in Fortran77 with quadruple precision for the
numerical calculation of the pentagons shown in Fig.\ref{fig-pen}.
The virtual EW correction to the \eezrr process can be expressed as
\begin{eqnarray}\label{eq-csvirt}
\Delta \sigma_{v}= \frac{1}{2!}\frac{(2\pi)^4}{2s}\int d\Phi_{3}
\overline{\sum_{spin}} 2Re\left\{{\cal M}_{v}{\cal
M}_{LO}^*\right\},
\end{eqnarray}
where ${\cal M}_{v}$ is the amplitude of all the virtual EW
correction Feynman diagrams.
%%%%%%%%%%%%%%%%%%%%%%%%%%%%%%

\par
\begin{figure*}
\begin{center}
\includegraphics [scale = 1.0]{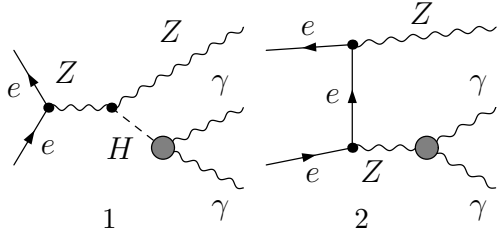}
\caption{\label{fig-res} The one-loop diagrams with possible
on-shell internal Higgs or $Z$-boson for the \eezrr process. }
\end{center}
\end{figure*}

The one-loop Feynman diagrams with possible Higgs and $Z$-boson
on-shell effects for the \eezrr process are shown in Fig.\ref{fig-res}.
Due to the Landau-Yang theorem \cite{Landau}, the contribution
from Fig.\ref{fig-res}(2) is vanished. The interference between the amplitude
of Fig.\ref{fig-res}(1) and the LO amplitude leads to a propagator factor of
$\frac{1}{(M_{\gamma\gamma}^2-M_{H}^2)}$, which is divergent in the
vicinity of $M_{\gamma\gamma}^2\sim M_{H}^2$. We regulate it by
making the replacement of $\frac{1}{(M_{\gamma\gamma}^2-M_{H}^2)} \to
\frac{1}{(M_{\gamma\gamma}^2-M_{H}^2 +iM_H\Gamma_H)}$.
We find that the contribution of this interference term is
so tiny that it can be ignored in the total NLO EW correction.

\par
The amplitude for all the one-loop Feynman diagrams contains both
the ultraviolet (UV) and the infrared (IR) singularities. We adopt
the dimensional regularization scheme \cite{DR}, in which the
dimensions of spinor and space-time manifolds are extended to
$D=4-2\epsilon$ to regularize the UV divergences in loop integrals,
and the IR singularities are regulated by adopting infinitesimal fictitious
photon mass as it commonly is applied to photon radiation in EW processes. The
relevant fields are renormalized by adopting the on-mass-shell (OMS)
renormalization scheme and the explicit expressions for the
renormalization constants are detailed in
Refs.\cite{ROSS,fourfermions}. As we expect, the UV divergence
contained in the loop virtual amplitude can be exactly canceled by that in
the counterterm amplitude.

\par
In order to get an IR-finite cross section for the \eezrr process at
the EW NLO, we consider the real photon emission process
$e^+(p_1)+e^-(p_2)\to Z(p_3)+\gamma(p_4)+\gamma(p_5)+\gamma(p_6)$.
The contribution of the real photon emission process has the form as
%%%%%%%%%%%%%%%%
\begin{eqnarray}\label{eq-csreal}
\Delta \sigma_{real}=\frac{1}{3!}\frac{(2\pi)^4}{2s}\int d\Phi_{4}
\overline{\sum_{spin}} |{\cal M}_{real}|^2,
\end{eqnarray}
%%%%%%%%%%%%%%%%%%%%%%
where $\frac{1}{3!}$ is due to the final three identical
photons. The phase space element of the four particles is defined as
\begin{eqnarray}
{d\Phi_4}=\delta^{(4)} \left( p_1+p_2-\sum_{i=3}^6p_i \right)
\prod_{j=3}^6 \frac{d^3 \vec{p}_j} {(2 \pi)^3 2 E_j}.
\end{eqnarray}

\par
By employing the dipole subtraction method we extract the IR singularities
from the real photon emission correction and combine them with the virtual
contribution. In this method the IR finite real correction is obtained by
subtracting an auxiliary function from the squared amplitude of the real
photon emission process before integrating over phase space due to the
subtraction function having the same singular structure as the squared
amplitude pointwise in phase space. The subtracted term is added again
after analytical integration over the
bremsstrahlung photon phase space. The dipole subtraction formalism
is a process independent approach which was first presented by Catani
and Seymour for QCD with massless partons \cite{catani1,catani2} and
subsequently was generalized to photon radiation off charged
particles with arbitrary mass by Dittmaier \cite{dps-ditt}. In our
calculations, we follow the approach of Ref.\cite{dps-ditt}, and we
check the independence on the parameter $\alpha \in (0,1]$
which essentially controls the region of dipole
phase space, such as $\alpha=1$ means the full dipole subtraction
being considered \cite{nagy}. Then the cancelation of IR singularities
is verified and the result shows that the NLO EW corrected cross
section for the \eezrr process is independent on the
IR regulator $m_{\gamma}$ in our calculation.

\par
To analyze the origin of the NLO EW corrections clearly, we
calculate the photonic (QED) and the genuine weak corrections
separately. The QED correction includes two parts: the QED virtual
correction $\Delta \sigma_{v}^{QED}$ which comes from the diagrams
with virtual photon exchange loop and the corresponding QED
parts of the counterterms, and the real photon emission correction
$\Delta \sigma_{real}$. The rest of the virtual electroweak
correction part is called the weak correction $\Delta
\sigma_{v}^{W}$. Therefore, the full NLO EW corrected cross section
can be expressed as
\begin{eqnarray}\label{totsigma}
\sigma_{NLO} &=& \sigma_{LO} + \Delta \sigma_{v}+\Delta \sigma_{real}
= \sigma_{LO} + \Delta \sigma_{v}^{QED}+\Delta \sigma_{v}^{W}+\Delta \sigma_{real} \nb \\
&=& \sigma_{LO} + \Delta \sigma^{QED}+\Delta \sigma_{v}^{W}
=\sigma_{LO}(1+\delta^{QED}+\delta^{W}) =
\sigma_{LO}(1+\delta^{EW}),
\end{eqnarray}
where the $\delta^{QED}$, $\delta^{W}$ and $\delta^{EW}$ are the
pure QED, genuine weak and full EW relative corrections,
respectively.

\vskip 5mm
\section{Numerical results and discussions}
\par
\subsection{Input parameters and kinematic cuts}
\par
For the numerical evaluation we adopt the $\alpha$-scheme and take
the following SM input parameters \cite{pdg}:
\begin{equation}\arraycolsep 2pt
\begin{array}[b]{lcllcllcl}
 M_{W} & = & 80.398~GeV,  &M_{Z}&  = & 91.1876~GeV,  &\Gamma_Z &= &2.4952~GeV \\
 m_e &=& 0.510998929~MeV, &m_{\mu}&=&105.6583715~MeV,&m_{\tau}&=&1.77682~GeV, \\
 m_u &=& 66~MeV, &m_c&=&1.275~GeV,&m_t&=&173.5~GeV, \\
 m_d &=& 66~MeV, &m_s&=&95~MeV,&m_b&=&4.65~GeV.
\end{array}
\label{SMpar}
\end{equation}
We take the fine structure constant $\alpha(0)=1/137.035999074$
defined in the Thomson limit. The current masses for light quarks
($m_u$ and $m_d$) can reproduce the hadronic contribution to the
shift in the fine structure constant $\alpha(M_Z)$
\cite{Jegerlehner}. We take the Higgs boson mass as $M_H=126~GeV$,
and its decay width is estimated by using the HDECAY program
\cite{hdecay}. The CKM matrix, whose matrix element appears only in
loop contribution, is set to be unity matrix.

\par
We apply the Cambridge/Aachen (C/A) jet algorithm
\cite{jet-algorithm} to photon candidates. For a three photon event
originating from the real emission correction, if the two final
photons with the smallest separation $R$ satisfy the constraint of
$R = \sqrt{\Delta y^2 + \Delta \phi^2} < 0.4$, where $\Delta y$ and
$\Delta \phi$ are the differences of rapidity and azimuthal angle
between the two photons, we combine this pair of photons as one new
photon track and this event is called as a 'two-photon' event
including the merged photon with four-momentum
$p_{ij,\mu}=p_{i,\mu}+p_{j,\mu}$, and contrariwise, it is called
as a 'three-photon' event. In our calculation we consider only
the 'two-photon' and 'three-photon' events with all the final photons
satisfying the constraints as
\begin{eqnarray}\label{cut}
p^{\gamma}_T \ge 15~GeV,~~~|y_{\gamma}|\le2.5,~~~
R_{\gamma\gamma}\ge 0.4.
\end{eqnarray}
Thereby we can exclude the inevitably infrared (IR) singularity at the tree level.
We name the photon in one event with the largest photon transverse energy as the leading photon,
while the photon with the next to largest photon transverse energy is named as the subleading photon.
In the 'inclusive' event selection scheme we collect all the 'two-photon' and 'three-photon'
events with the limitations on photons shown in Eq.(\ref{cut}). In the 'exclusive' event
selection scheme, we include only the so-called 'two-photon' events satisfying the
constraints as shown in Eq.(\ref{cut}). In following discussion we adopt the 'inclusive'
scheme for event selection as default unless otherwise stated.

\par
\subsection{Total cross section}
\par
\begin{figure}[htbp]
\includegraphics[angle=0,width=3.2in,height=2.4in]{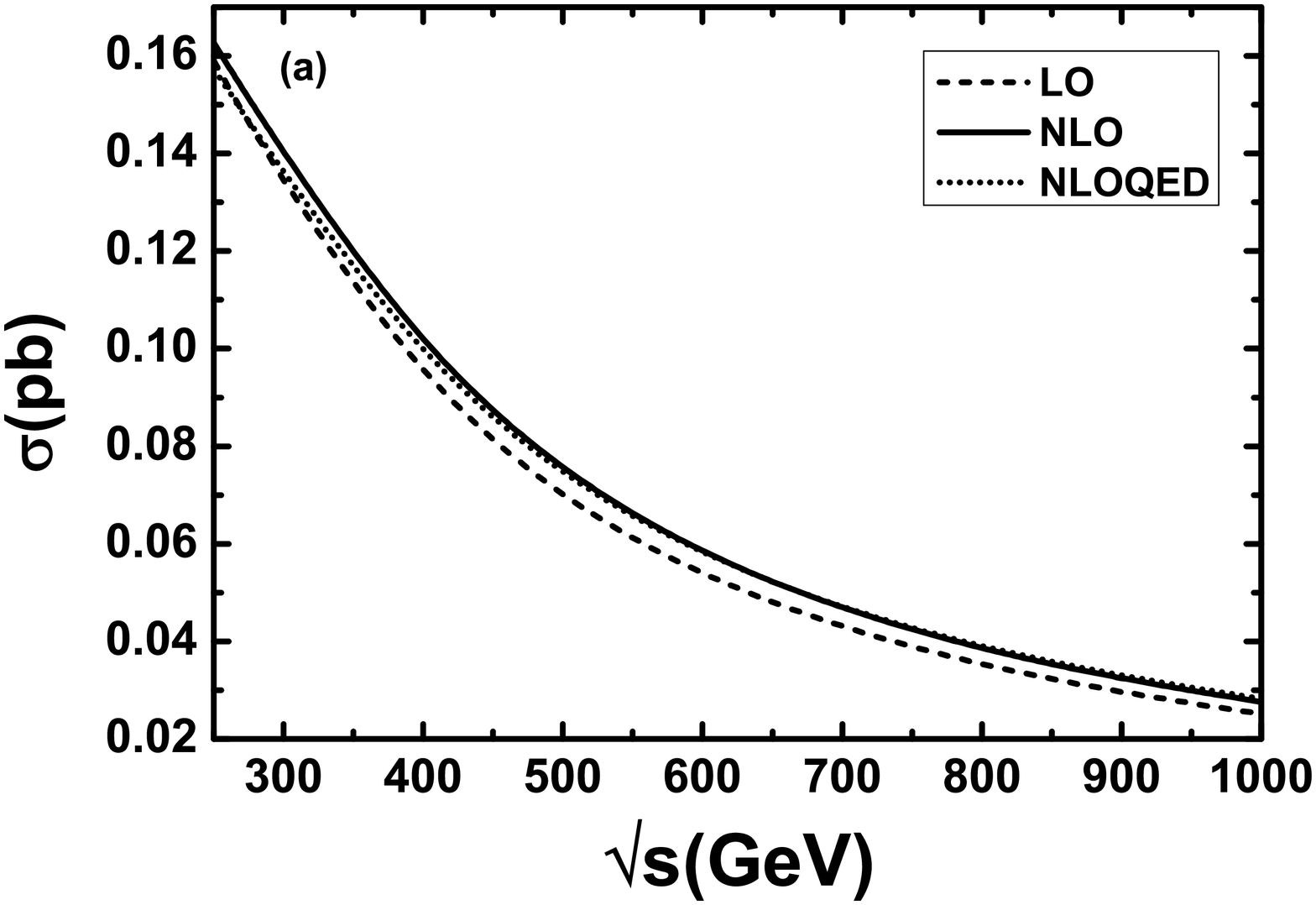}%
\hspace{0in}%
\includegraphics[angle=0,width=3.2in,height=2.4in]{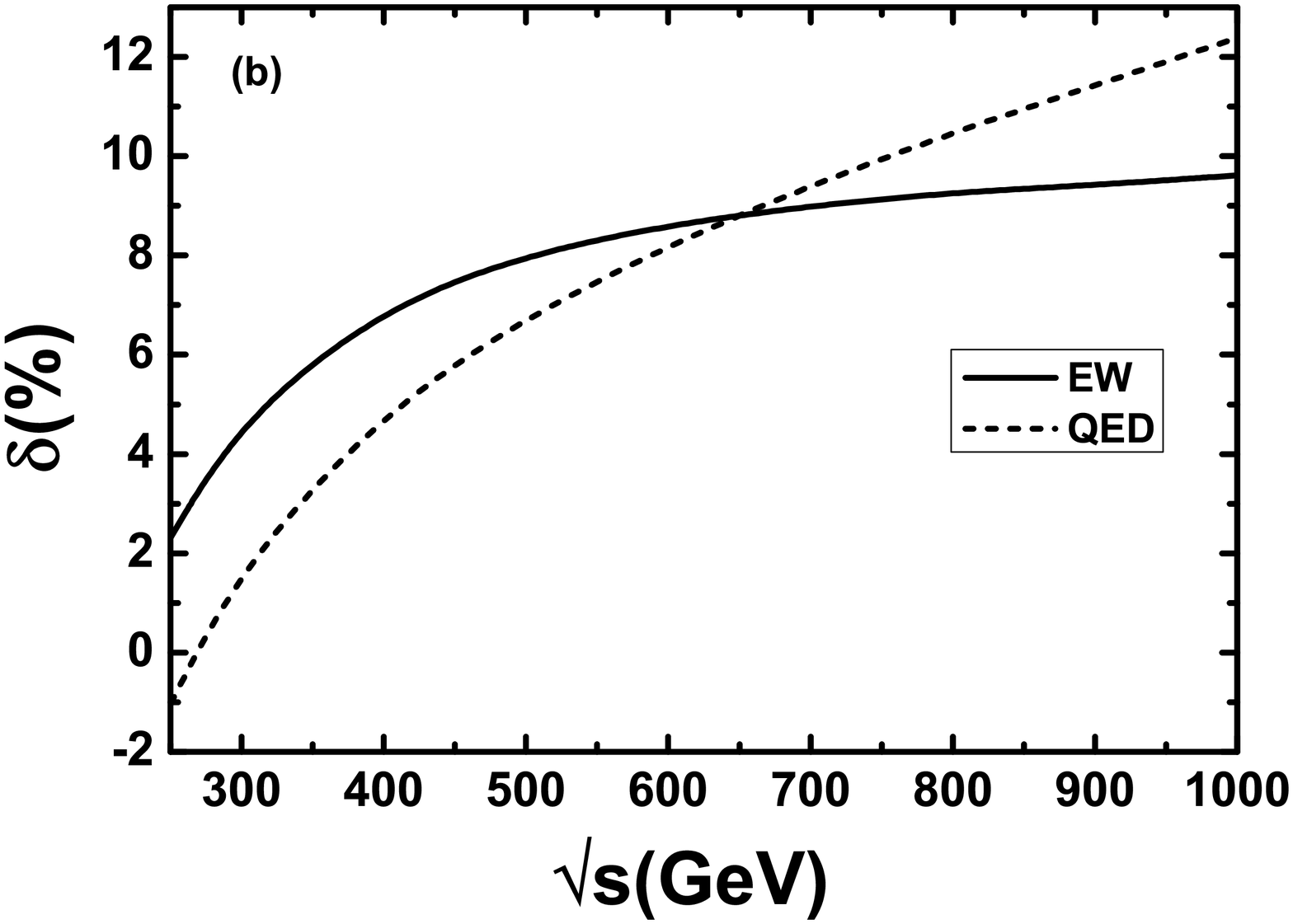}%
\hspace{0in}%
\caption{(a) The LO, NLO EW and the pure NLO QED corrected cross
sections ($\sigma_{LO}$, $\sigma_{NLO}$, $\sigma_{NLO}^{QED}$) for
the \eezrr process as the functions of the colliding energy
$\sqrt{s}$ in the 'inclusive' event selection scheme at the ILC. (b)
The corresponding NLO EW and pure NLO QED relative corrections
($\delta^{EW}$, $\delta^{QED}$).} \label{fig-sqrts}
\end{figure}

The dependence of the LO integrated cross section for the \eezrr
process in the SM on the colliding energy was presented in Fig.1 of
Ref.\cite{Stirling:2000}. When we take the same input parameters as
in that reference, the coincident numerical results can be obtained.
In Fig.\ref{fig-sqrts}(a), we plot the LO, NLO EW and pure NLO QED
corrected integrated cross sections as the functions of the
colliding energy $\sqrt{s}$ in the 'inclusive' event selection
scheme, and in Fig.\ref{fig-sqrts}(b) we show the corresponding NLO
EW and pure NLO QED relative corrections, $\delta^{EW}\equiv
\frac{\sigma_{NLO}-\sigma_{LO}}{\sigma_{LO}}$ and
$\delta^{QED}\equiv
\frac{\sigma_{NLO}^{QED}-\sigma_{LO}}{\sigma_{LO}}$. Some
representative numerical results read out from
Figs.\ref{fig-sqrts}(a) and (b) are listed in Table \ref{tab1}. From
these figures we find all the curves for the cross sections decrease
quickly with the increment of $\sqrt s$, and the LO cross section is
always enhanced by the NLO EW correction in the whole $\sqrt s$
plotted range. When $\sqrt s$ goes up from $250~GeV$ to $1~TeV$, the
NLO EW relative correction $\delta^{EW}$ varies from $2.32\%$ to
$9.61\%$. We also see that the pure NLO QED correction part always
increases the LO cross section when $\sqrt s> 270~GeV$ and the pure
NLO QED relative correction becomes more and more notable with the
increment of $\sqrt s$. In order to make a comparison of the results
in different event selection schemes, we also present
corresponding numerical results by adopting the 'exclusive' event
selection scheme in Table \ref{tab2}. We can see that with the same
$\sqrt s$ the NLO EW and pure NLO QED corrected cross sections in
Table \ref{tab2} are less than the corresponding ones by adopting
the 'inclusive' event selection scheme, due to that all the so-called
'three-photon' events are abandoned in the 'exclusive' event selection scheme.
%%%%%%%%%%%%%%%%%%%%
\begin{table}
\center
\begin{tabular}{cccccc}
\hline
$\sqrt{s}(GeV)$ & $\sigma_{LO}(fb)$ & $\sigma_{NLO}(fb)$
&$\sigma_{NLO}^{QED}(fb)$ & $\delta^{EW}(\%)$& $\delta^{QED}(\%)$ \\
\hline \hline
250   & 159.05(4)  & 162.73(13)   &157.39(18)    & 2.32    &-1.04  \\
300   & 133.41(4)  & 139.71(12)   &135.79(17)    & 4.72    & 1.79\\
400   & 93.12(3)   & 99.61(7)     &97.65(9)      & 6.97    & 4.86\\
500   & 68.74(2)   & 74.25(4)     &73.40(5)      & 8.02    & 6.77\\
600   & 53.18(2)   & 57.76(5)     &57.54(7)      & 8.61    & 8.20\\
700   & 42.62(1)   & 46.46(3)     &46.63(4)      & 9.00    & 9.42\\
800   & 35.07(1)   & 38.32(3)     &38.75(3)      & 9.27    & 10.48\\
900   & 29.47(1)   & 32.25(3)     &32.84(3)      & 9.42    & 11.42\\
1000  & 25.12(1)   & 27.53(3)     &28.23(3)      & 9.61    & 12.39\\
\hline  \hline
\end{tabular}
\caption{ \label{tab1} The total LO, NLO EW, pure NLO QED corrected
integrated cross sections ($\sigma_{LO}$, $\sigma_{NLO}$ and
$\sigma^{QED}_{NLO}$), and the corresponding EW and QED relative
corrections ($\delta^{EW}$ and $\delta^{QED}$)
for the $e^+ e^- \to Z \gamma \gamma$ process in the 'inclusive' event
selection scheme.}
\end{table}
%%%%%%%%%%%%%%%%%%%%
\begin{table}
\center
\begin{tabular}{cccccc}
\hline
$\sqrt{s}(GeV)$ & $\sigma_{LO}(fb)$ & $\sigma_{NLO}(fb)$
&$\sigma_{NLO}^{QED}(fb)$ & $\delta^{EW}(\%)$& $\delta^{QED}(\%)$ \\
\hline \hline
250   & 159.05(4)  & 161.81(13)   &156.47(14)    & 1.74    &-1.62  \\
300   & 133.41(4)  & 138.56(12)   &134.64(12)    & 3.86    & 0.92\\
400   & 93.12(3)   & 98.39(7)     &96.43(9)      & 5.66    & 3.55\\
500   & 68.74(2)   & 73.12(4)     &72.26(5)      & 6.37    & 5.12\\
600   & 53.18(2)   & 56.74(5)     &56.52(7)      & 6.69    & 6.28\\
700   & 42.62(1)   & 45.53(3)     &45.71(4)      & 6.83    & 7.24\\
800   & 35.07(1)   & 37.48(3)     &37.91(3)      & 6.88    & 8.09\\
900   & 29.47(1)   & 31.49(3)     &32.08(3)      & 6.85    & 8.85\\
1000  & 25.12(1)   & 26.85(3)     &27.54(3)      & 6.87    & 9.64\\
\hline  \hline
\end{tabular}
\caption{ \label{tab2}  The total LO cross section ($\sigma_{LO}$), NLO
EW, pure NLO QED corrected integrated cross sections ($\sigma_{NLO}$
and $\sigma^{QED}_{NLO}$), and the corresponding EW and QED relative
corrections ($\delta^{EW}$ and $\delta^{QED}$)
for the $e^+ e^- \to Z \gamma \gamma$ process in the 'exclusive' event
selection scheme.}
\end{table}
%%%%%%%%%%%%%%%%%
%%%%%%%%%%%%%%%%%%

\par
\subsection{Kinematic distributions}
\par
\begin{figure}[htbp]
\includegraphics[angle=0,width=3.2in,height=2.4in]{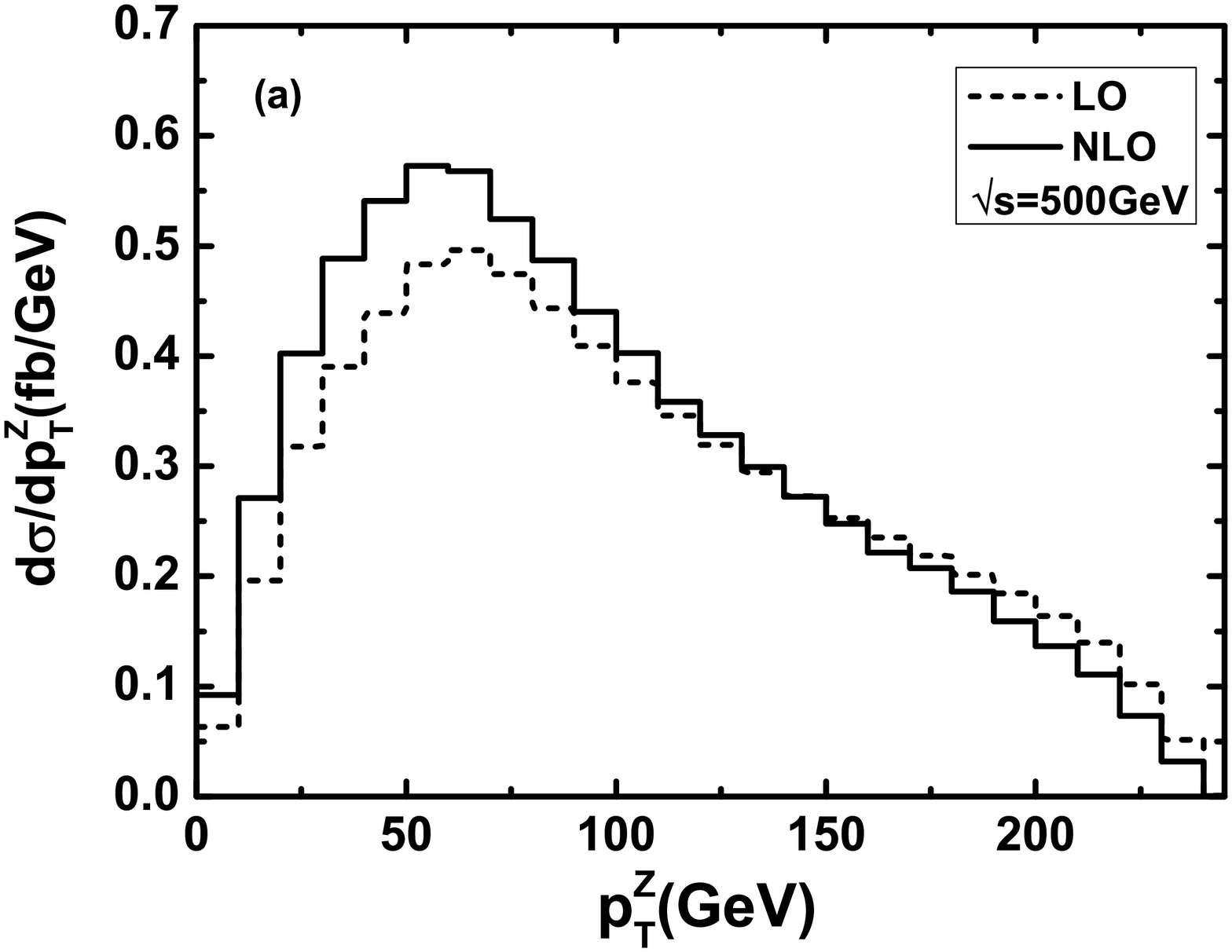}%
\hspace{0in}%
\includegraphics[angle=0,width=3.2in,height=2.4in]{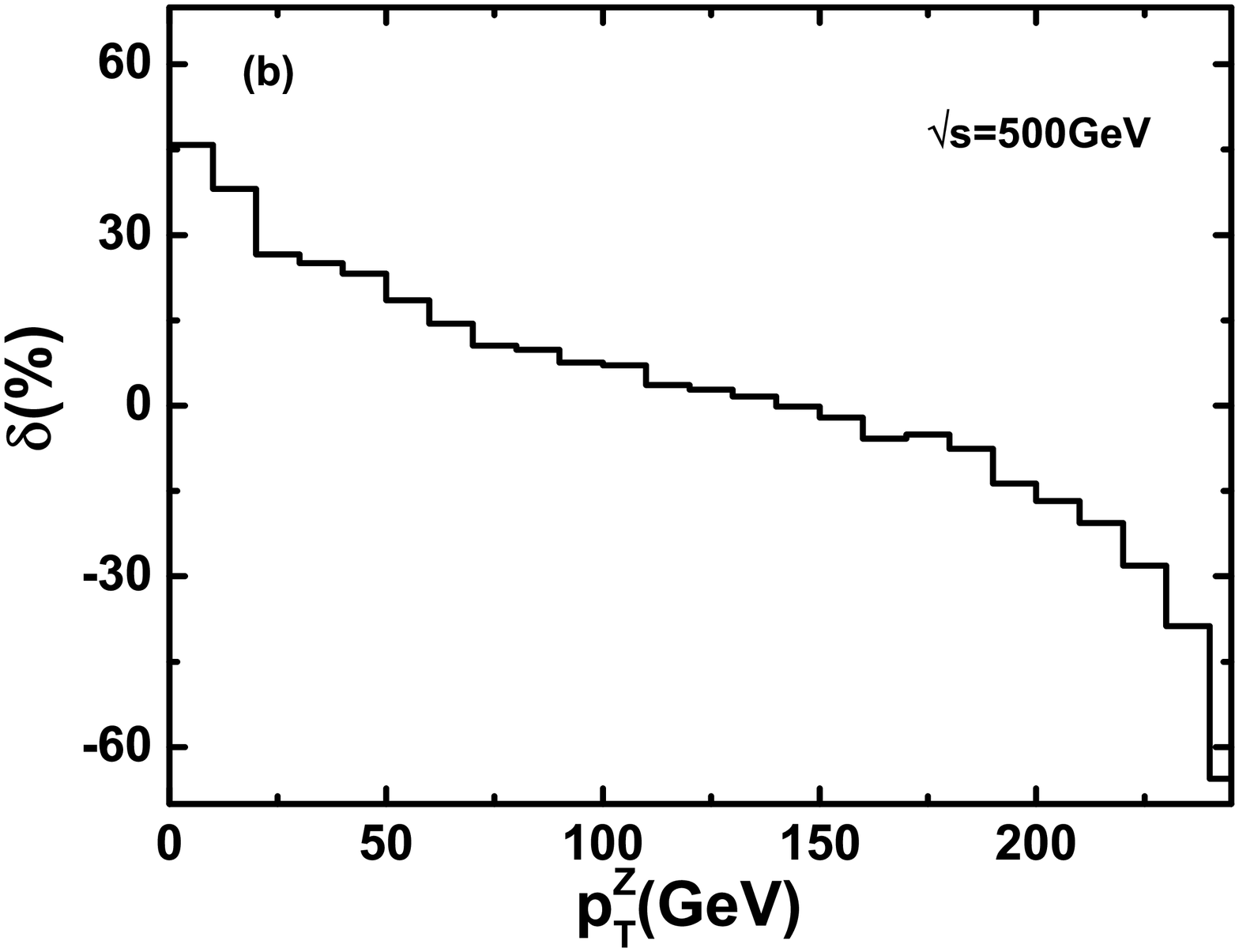}%
\hspace{0in}%
\caption{ (a) The LO and NLO EW corrected transverse momentum
distributions of $Z$-boson with $\sqrt s =500~GeV$ in the 'inclusive'
event selection scheme. (b) The corresponding NLO EW relative
corrections.  } \label{fig-ptz}
\end{figure}
\begin{figure}[htbp]
\includegraphics[angle=0,width=3.2in,height=2.4in]{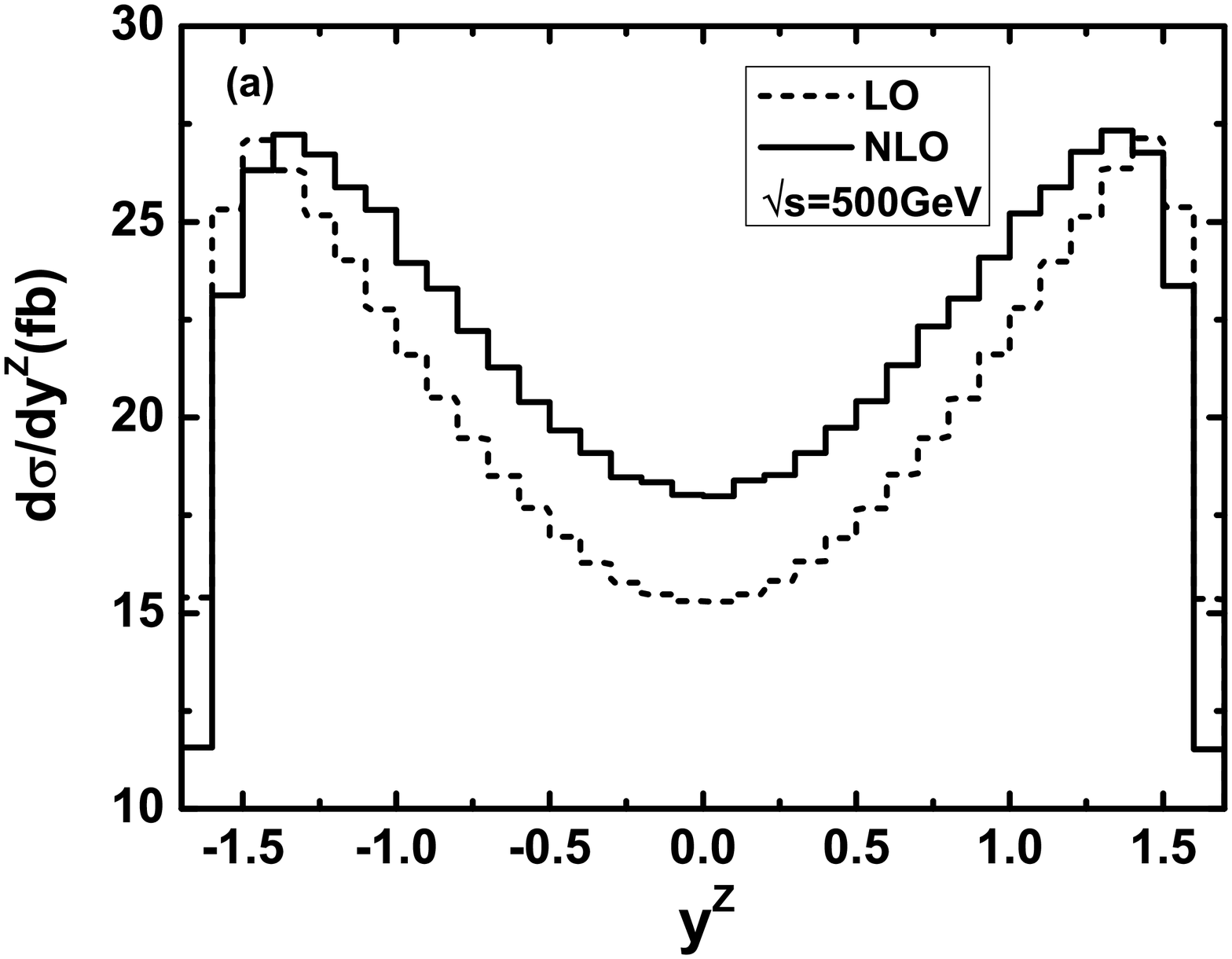}%
\hspace{0in}%
\includegraphics[angle=0,width=3.2in,height=2.4in]{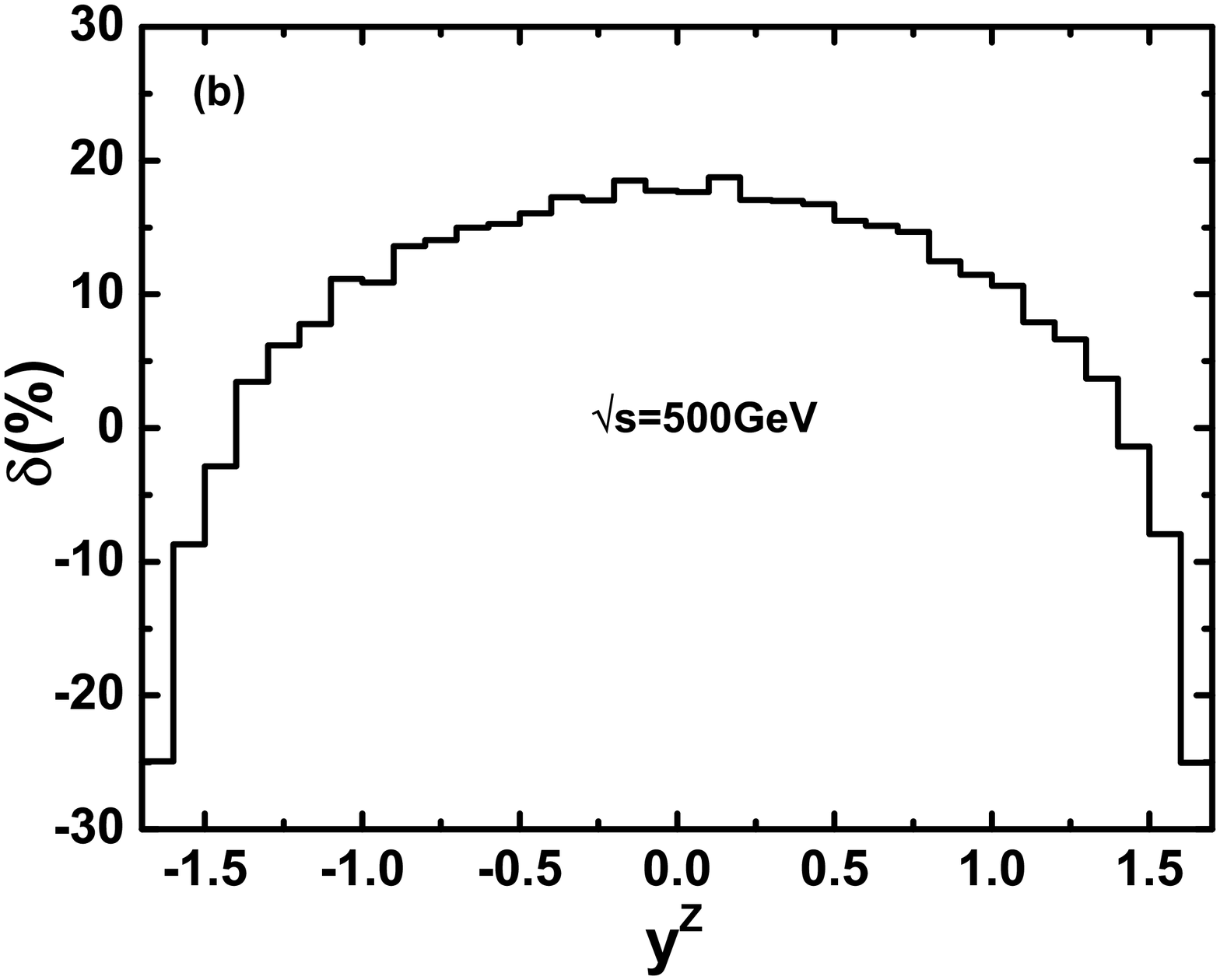}%
\hspace{0in}%
\caption{(a) The LO and NLO EW corrected rapidity distributions of
$Z$-boson with $\sqrt s =500~GeV$ in the 'inclusive' event selection
scheme. (b) The corresponding NLO EW relative corrections. }
\label{fig-rapz}
\end{figure}

We present the LO and NLO EW corrected transverse momentum and
rapidity distributions of the final $Z$-boson in
Fig.\ref{fig-ptz}(a) and Fig.\ref{fig-rapz}(a), respectively. The
corresponding EW relative corrections $\delta^{EW}$ are also plotted
in Fig.\ref{fig-ptz}(b) and Fig.\ref{fig-rapz}(b), separately. There
the results are obtained by taking $\sqrt s = 500~GeV$ and applying
the 'inclusive' event selection scheme. From Figs.\ref{fig-ptz}(a,b)
we can see that the NLO EW correction enhances the LO differential
cross section $d\sigma_{LO}/dp_T^Z$ in low $p_T^Z$ region. The NLO
relative correction always goes down with the increment of $p_T^Z$,
and changes from being positive to negative when $p_T^Z$ arrives at
the position about $145~GeV$. In Figs.\ref{fig-rapz}(a,b), we find
that the LO rapidity distribution is strengthened obviously by the
NLO EW correction in the central rapidity region of $Z$-boson at the
ILC, while weakened by the quantum correction in the regions of
$|y_Z|\ge 1.4$.
%%%%%%%%%%%%%%%%%%
%%%%%%%%%%%%%%%%%%%

\par
\begin{figure}[htbp]
\includegraphics[angle=0,width=3.2in,height=2.4in]{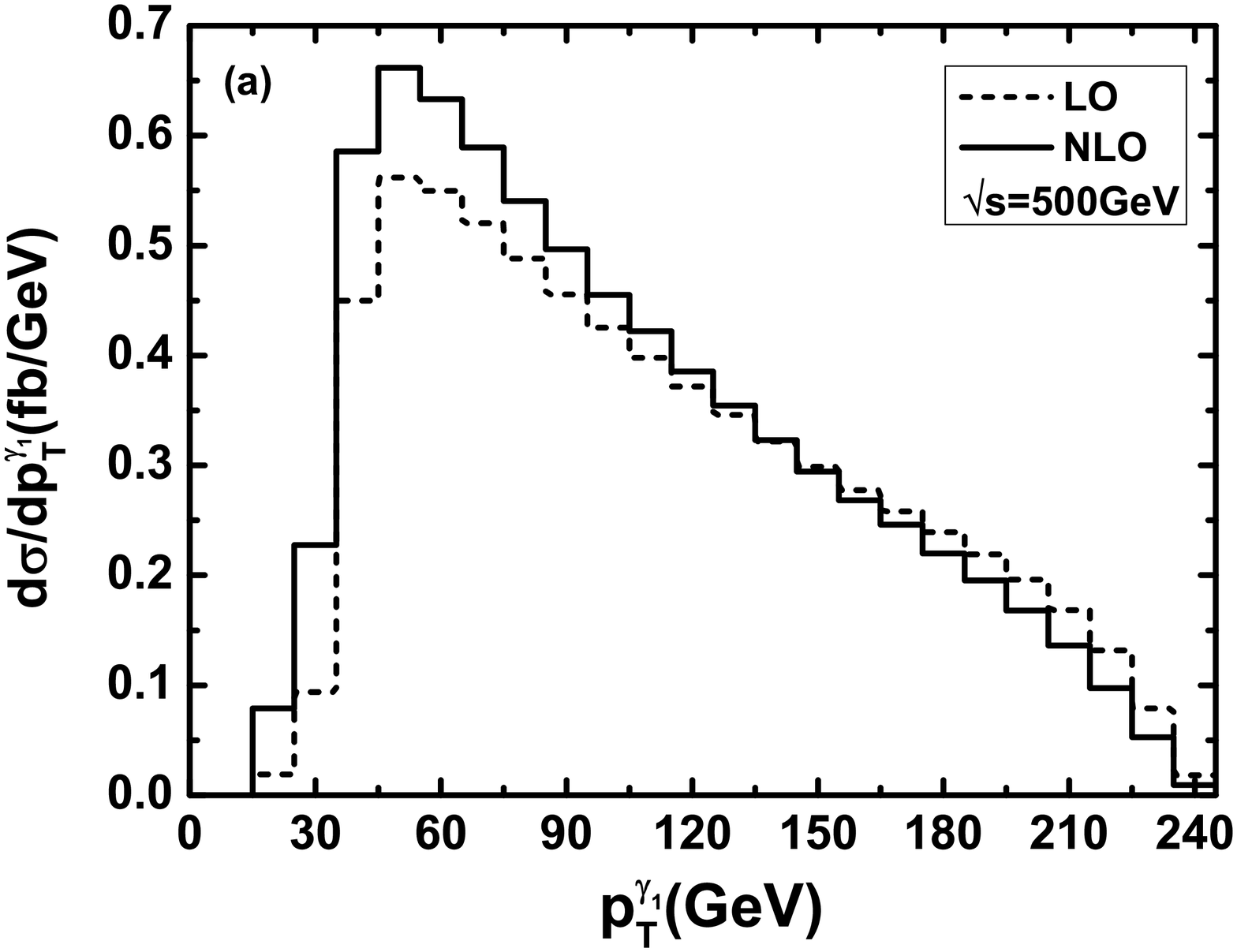}%
\hspace{0in}%
\includegraphics[angle=0,width=3.2in,height=2.4in]{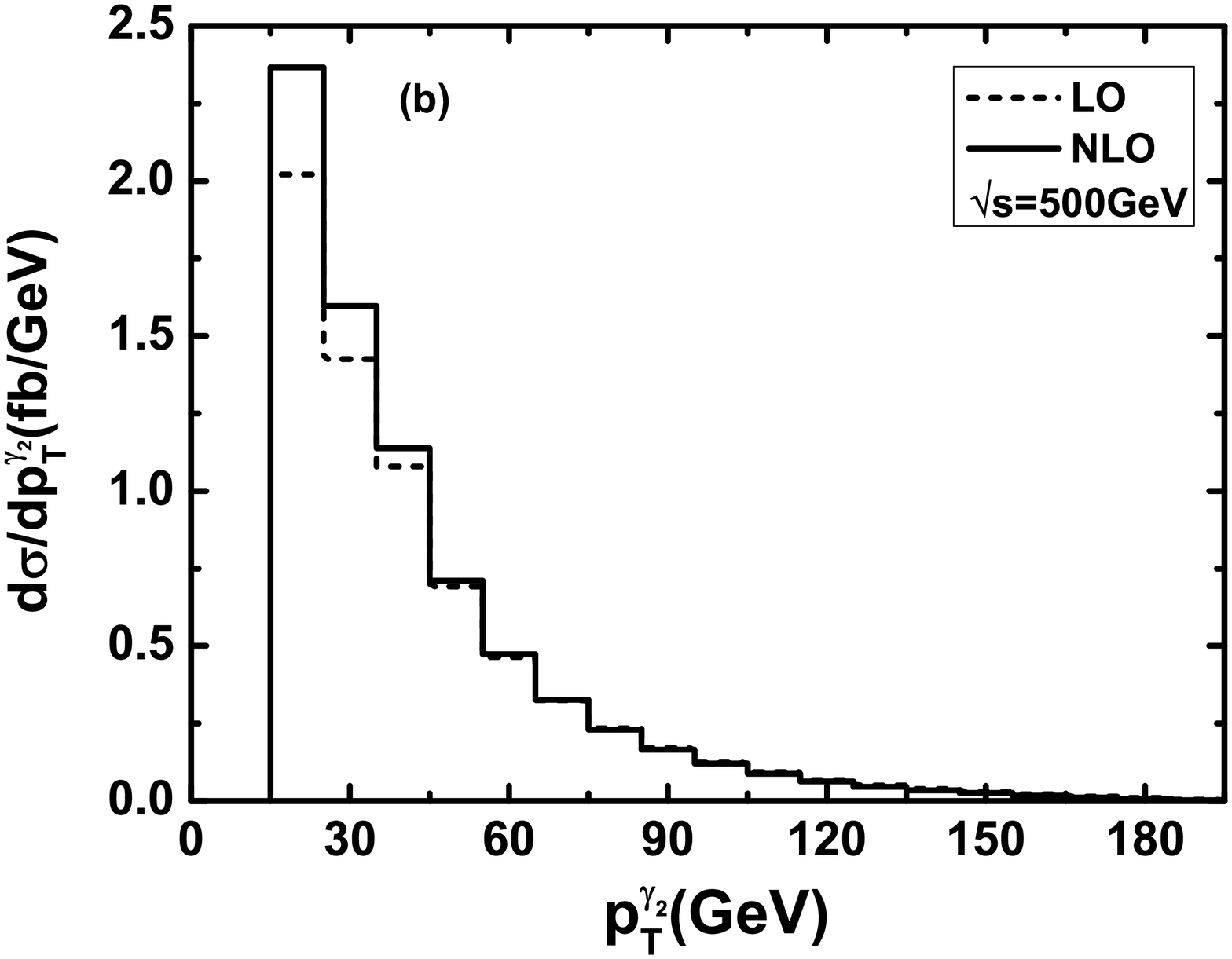}%
\hspace{0in}%
\caption{The LO and NLO EW corrected transverse momentum distributions of the
final photons with $\sqrt s =500~GeV$ in the 'inclusive' event selection scheme.
(a) For the leading photon. (b) For the subleading photon.  }
\label{fig-ptr}
\end{figure}
\begin{figure}[htbp]
\includegraphics[angle=0,width=3.2in,height=2.4in]{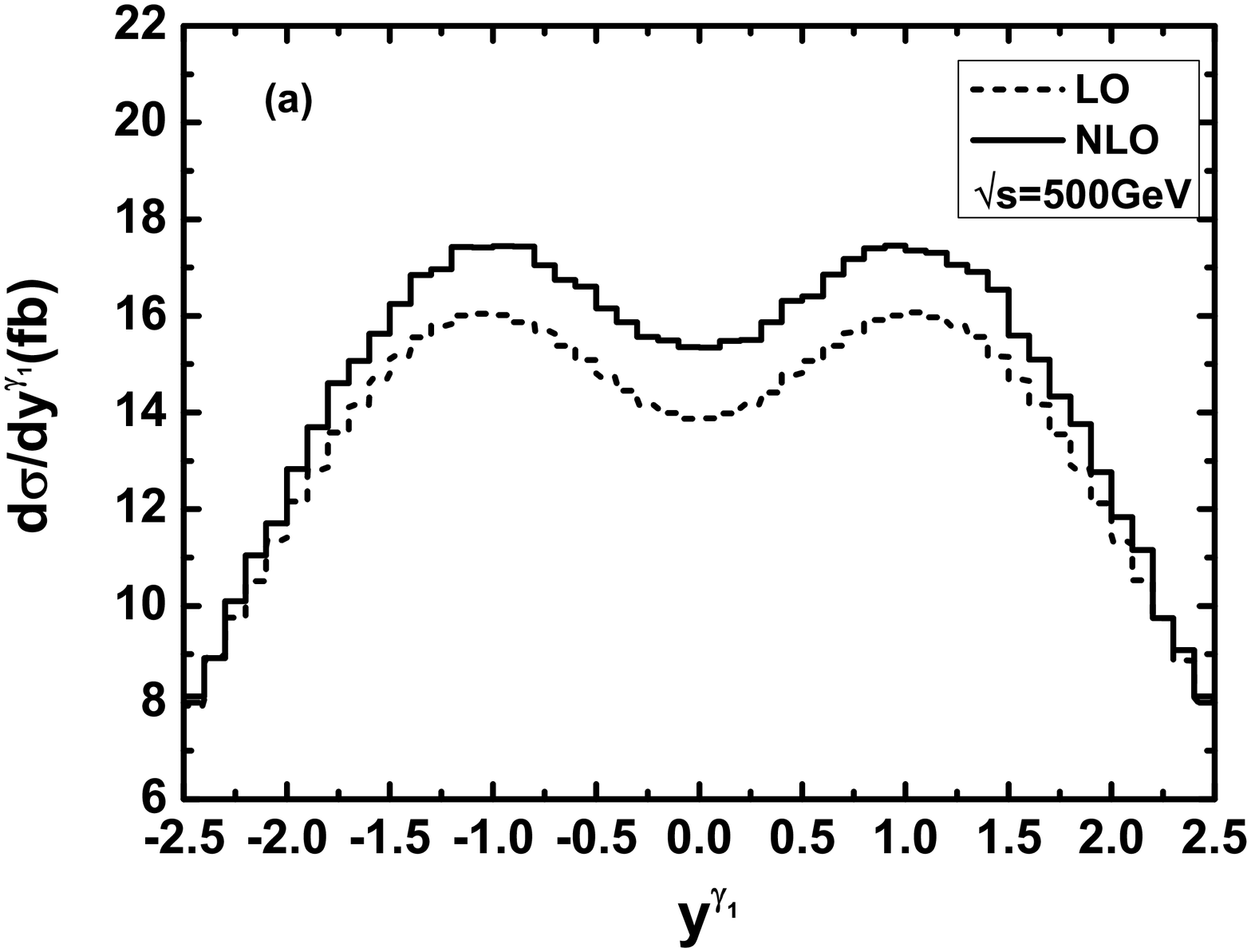}%
\hspace{0in}%
\includegraphics[angle=0,width=3.2in,height=2.4in]{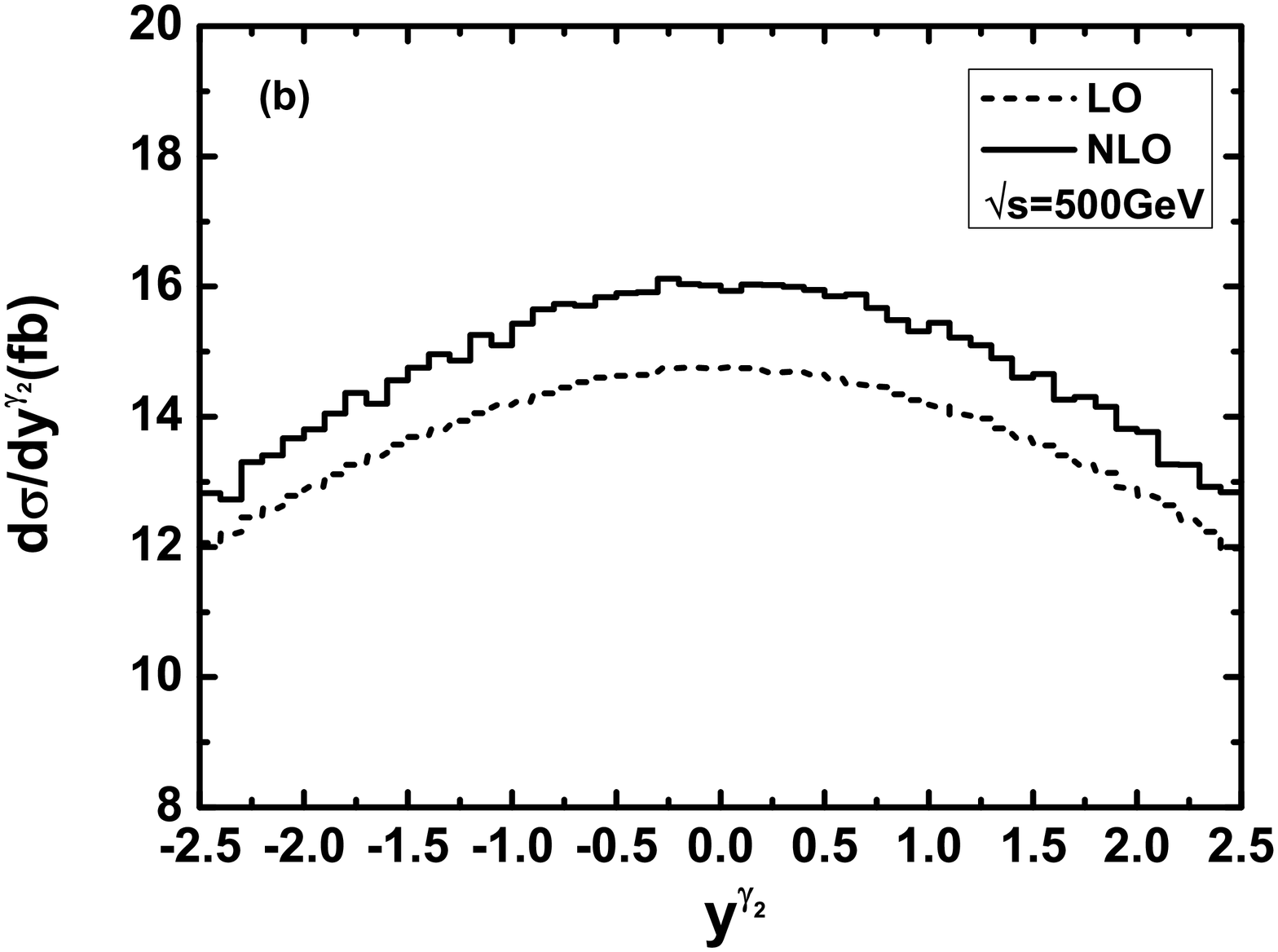}%
\hspace{0in}%
\caption{ The LO and NLO EW corrected rapidity distributions of the
final photons with $\sqrt s =500~GeV$ in the 'inclusive' event selection scheme.
(a) For the leading photon. (b) For the subleading photon.  } \label{fig-rapr}
\end{figure}

The transverse momentum distributions of the leading photon (labeled
by $\gamma_1$) and the subleading photon (labeled by $\gamma_2$) are
plotted in Fig.\ref{fig-ptr}(a) and  Fig.\ref{fig-ptr}(b),
respectively. The rapidity distributions of the leading and
subleading photons are presented in Fig.\ref{fig-rapr}(a) and
Fig.\ref{fig-rapr}(b), separately. In these four figures we adopt
the 'inclusive' event selection scheme, and we take $\sqrt s =500~GeV$,
the cuts on photons being declared in Eq.(\ref{cut}). It can be seen from both
Figs.\ref{fig-ptr}(a) and (b) that the LO $p_T$ distributions for
the leading and subleading photons are enhanced in the lower $p_T$
region (i.e., $p_T^{\gamma_1} < 145~GeV$ and $p_T^{\gamma_2} <
75~GeV$, separately), but they are suppressed in the rest of $p_T$ regions by
the NLO corrections. The LO and NLO EW corrected transverse momentum
distributions for the leading photon reach their maxima at the
position about $50~GeV$. While for the subleading photon, the LO and
NLO EW corrected transverse momentum distributions always decrease
with the increment of $p_T$. Figs.\ref{fig-rapr}(a,b) show that the
rapidity distributions of the leading and subleading photons are
both reinforced by the NLO EW corrections in the whole plotted
rapidity region. From the figures we see that both the LO and NLO
corrected rapidity distributions for the leading photon have two
peaks, which are located at the positions of $|y|\sim 1$, in
contrast the subleading photon rapidity distributions reach their
maxima in the central rapidity region.
%%%%%%%%%%%%%%%%%%%%%
%%%%%%%%%%%%%%%%%%

\par
\begin{figure}[htbp]
\includegraphics[angle=0,width=3.2in,height=2.4in]{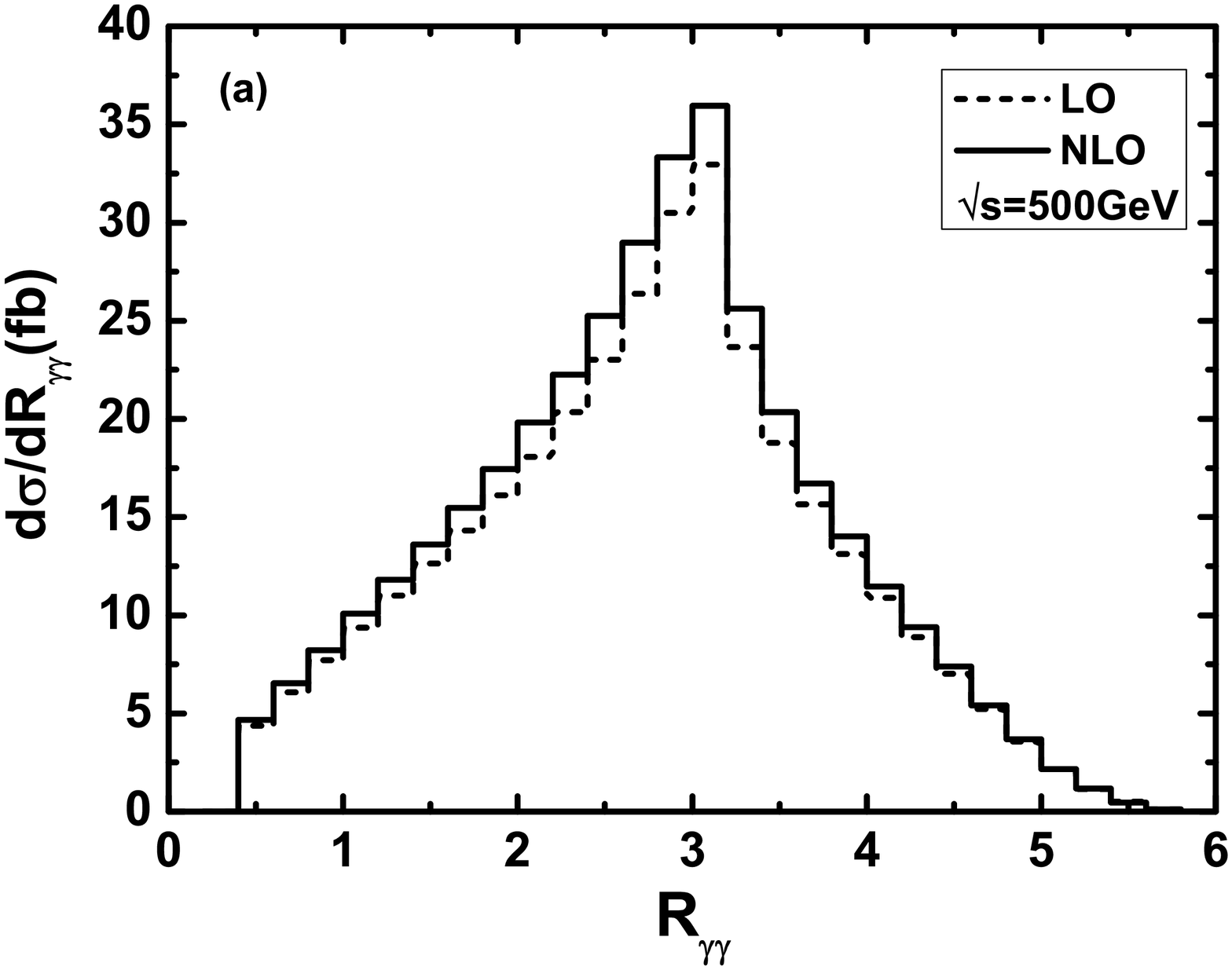}%
\hspace{0in}%
\includegraphics[angle=0,width=3.2in,height=2.4in]{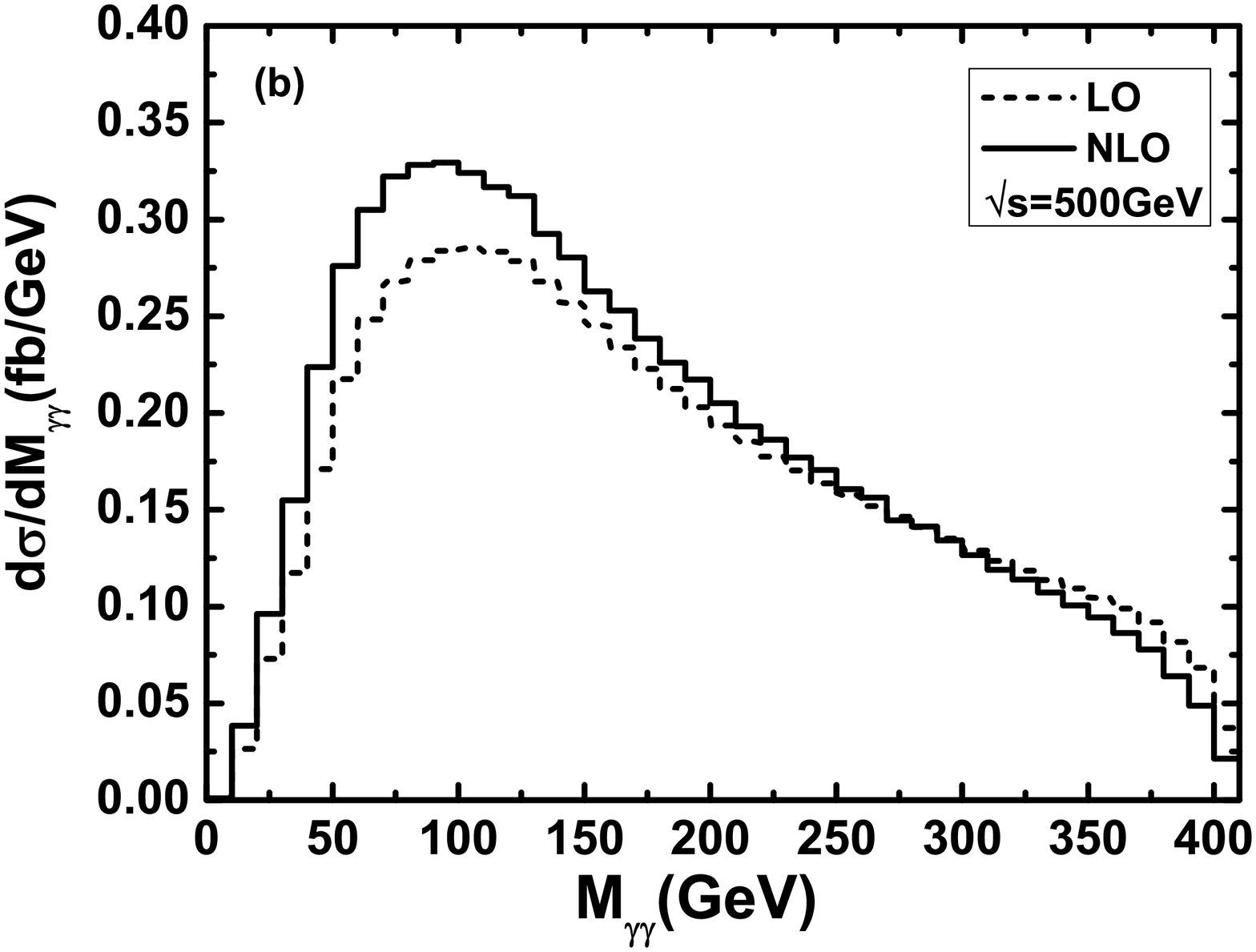}%
\hspace{0in}%
\caption{(a) The LO and NLO EW corrected distributions of the
separation $R_{\gamma\gamma}$ between the final leading and subleading photons.
(b) The LO and NLO EW corrected invariant mass $M_{\gamma\gamma}$
distributions. } \label{fig-mrr}
\end{figure}

The LO and NLO EW corrected distributions of the separation
$R_{\gamma \gamma}$ between the final leading and subleading photons
are plotted in Fig.\ref{fig-mrr}(a). It shows that at both the LO
and the NLO the preferred kinematical configuration of the leading
and subleading photons wide separation in the rapidity-azimuthal-angle
plane, and the LO and NLO
$R_{\gamma\gamma}$ distributions reach their maxima at the location
of $R_{\gamma \gamma} \sim 3$. In Fig.\ref{fig-mrr}(b), we depict
the LO and NLO EW corrected distributions of the invariant mass of
the leading and subleading photons (denoted as $M_{\gamma\gamma}$).
It demonstrates that both the LO and NLO EW corrected
$M_{\gamma\gamma}$ distributions reach their maxima in the vicinity
of $M_{\gamma\gamma}\sim 100~GeV$, and the NLO EW correction
enhances the LO differential cross section in the range of
$M_{\gamma\gamma}\le 270~GeV$.

\par
From the figures of Figs.\ref{fig-ptz}-\ref{fig-mrr} we can see that
the phase space dependence of the NLO EW correction is nontrivial
and sizable, and the NLO EW correction does not observably change
the LO distribution line shape in the case of taking the 'inclusive'
event selection scheme.
%%%%%%%%%%%%%%%%%%

\par
\begin{figure}[htbp]
\includegraphics[angle=0,width=3.2in,height=2.4in]{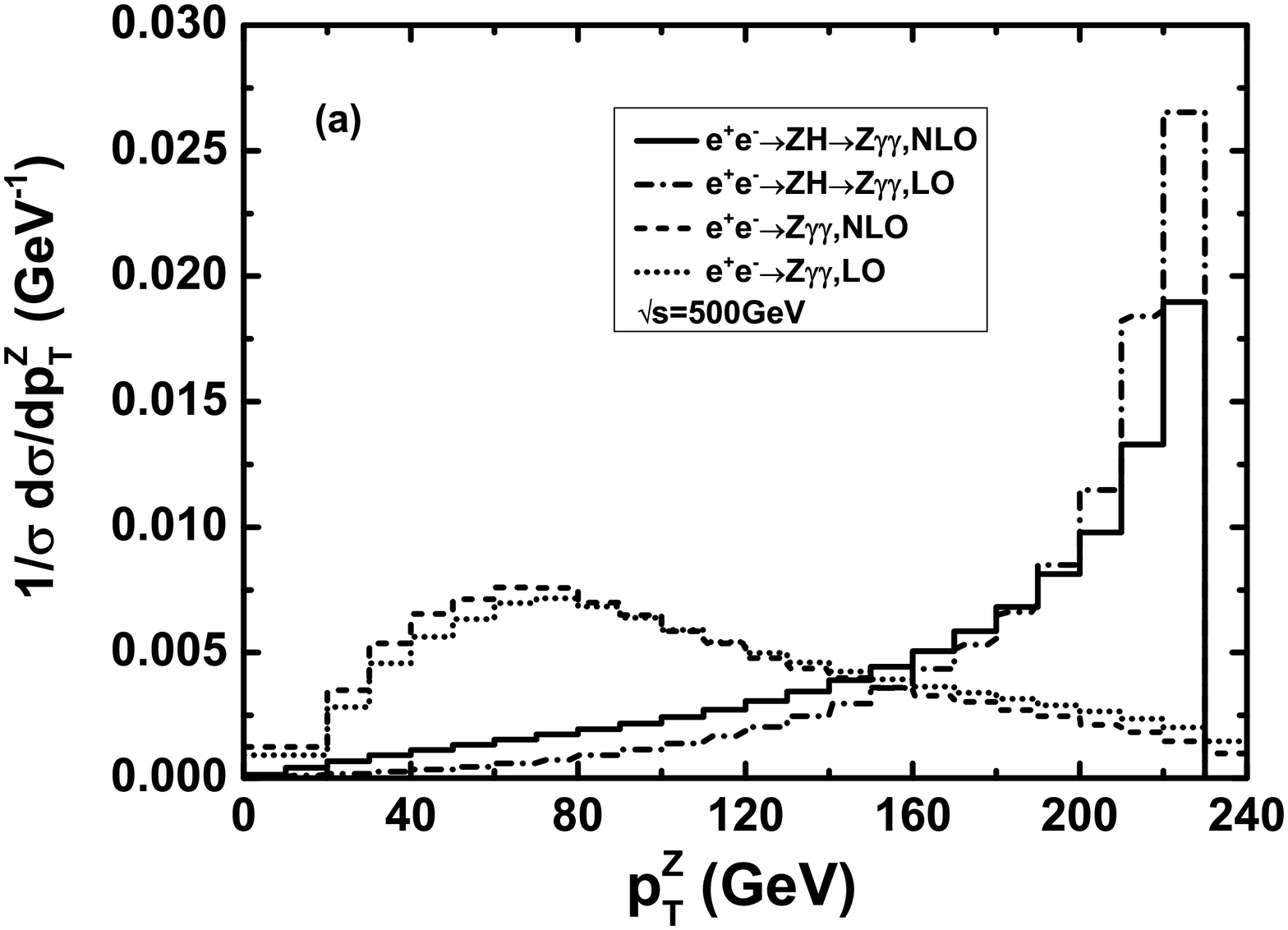}%
\hspace{0in}%
\includegraphics[angle=0,width=3.2in,height=2.4in]{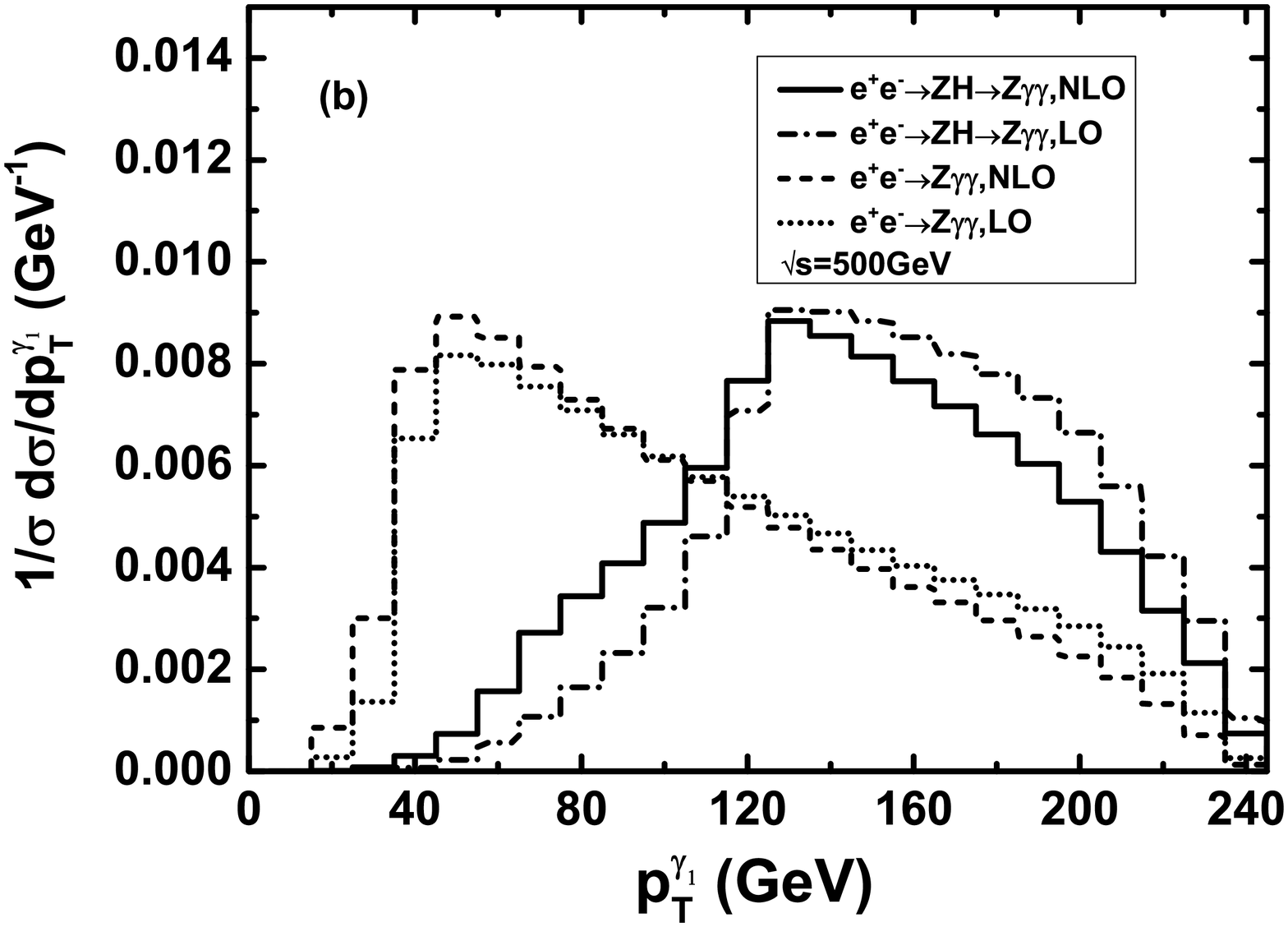}%
\hspace{0in}%
\includegraphics[angle=0,width=3.2in,height=2.4in]{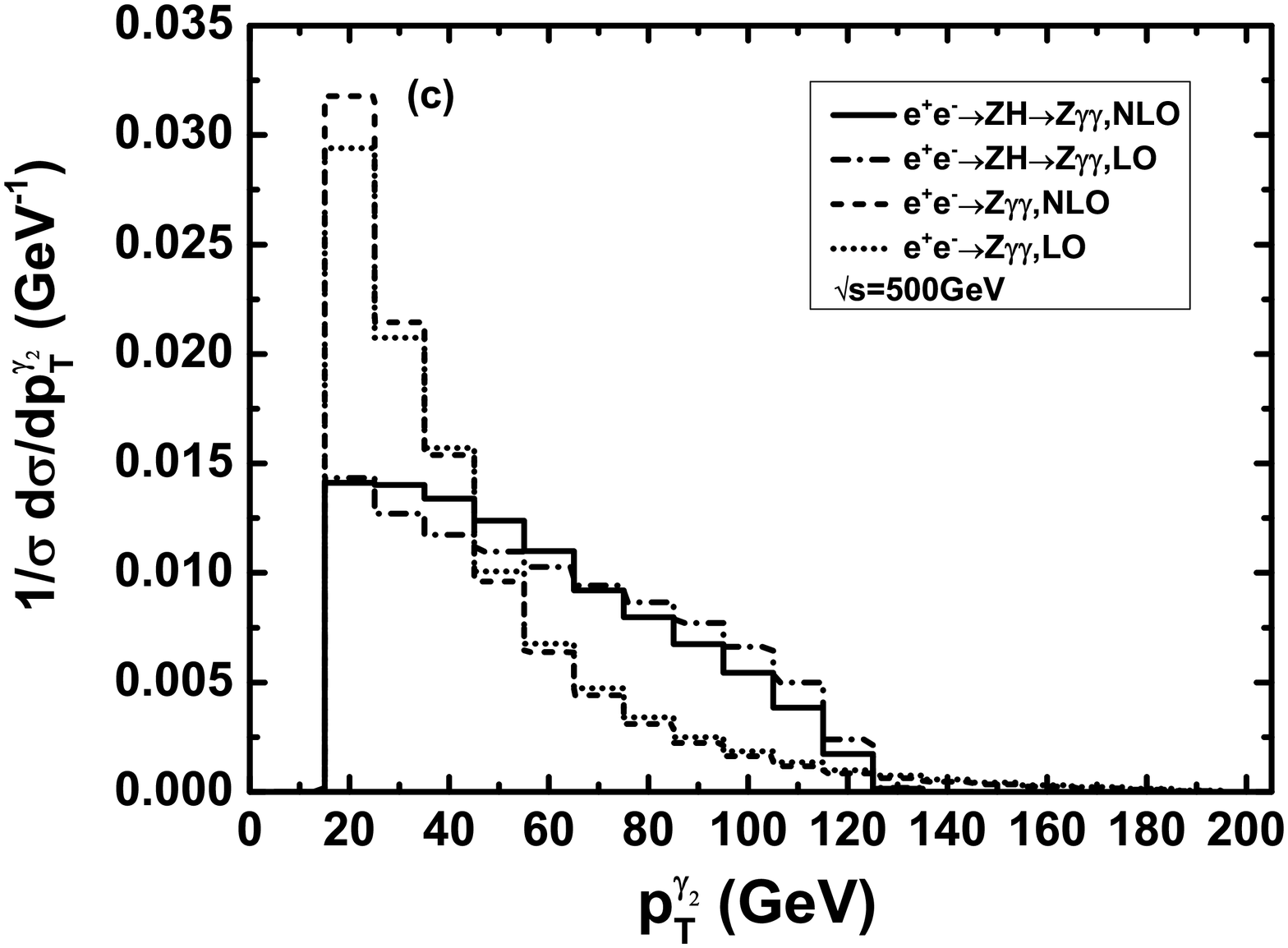}%
\hspace{0in}%
\includegraphics[angle=0,width=3.2in,height=2.4in]{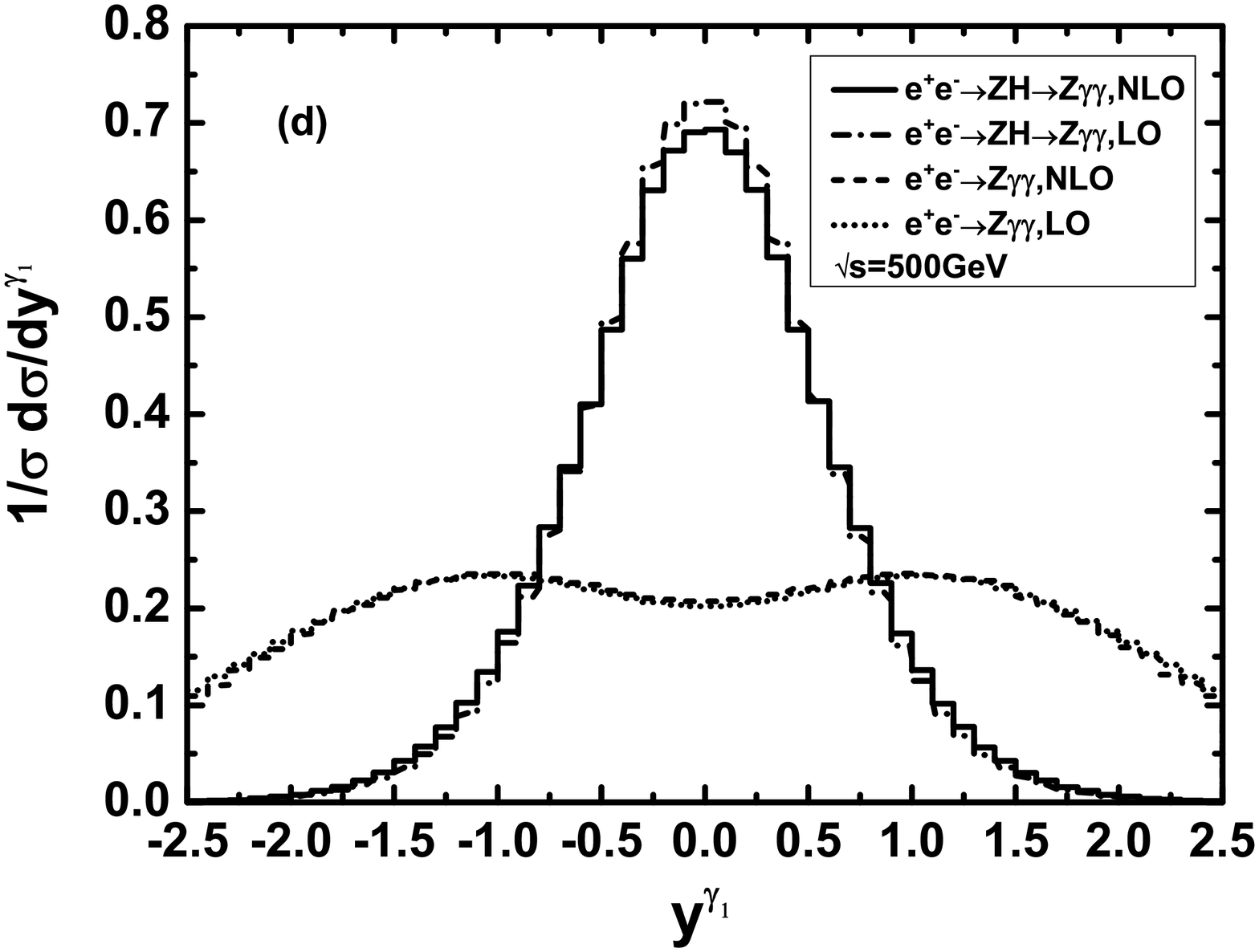}%
\hspace{0in}%
\includegraphics[angle=0,width=3.2in,height=2.4in]{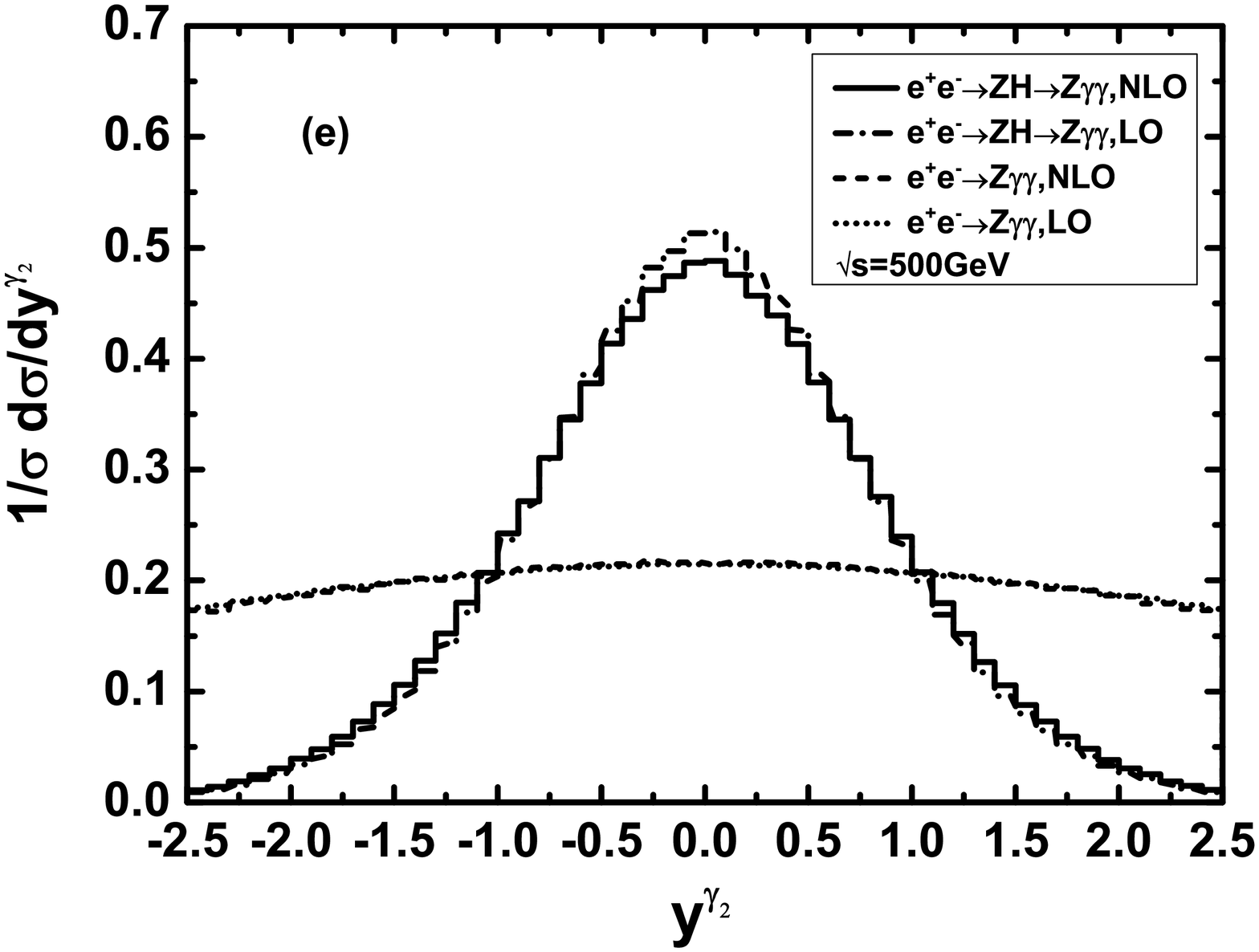}%
\hspace{0in}%
\includegraphics[angle=0,width=3.2in,height=2.4in]{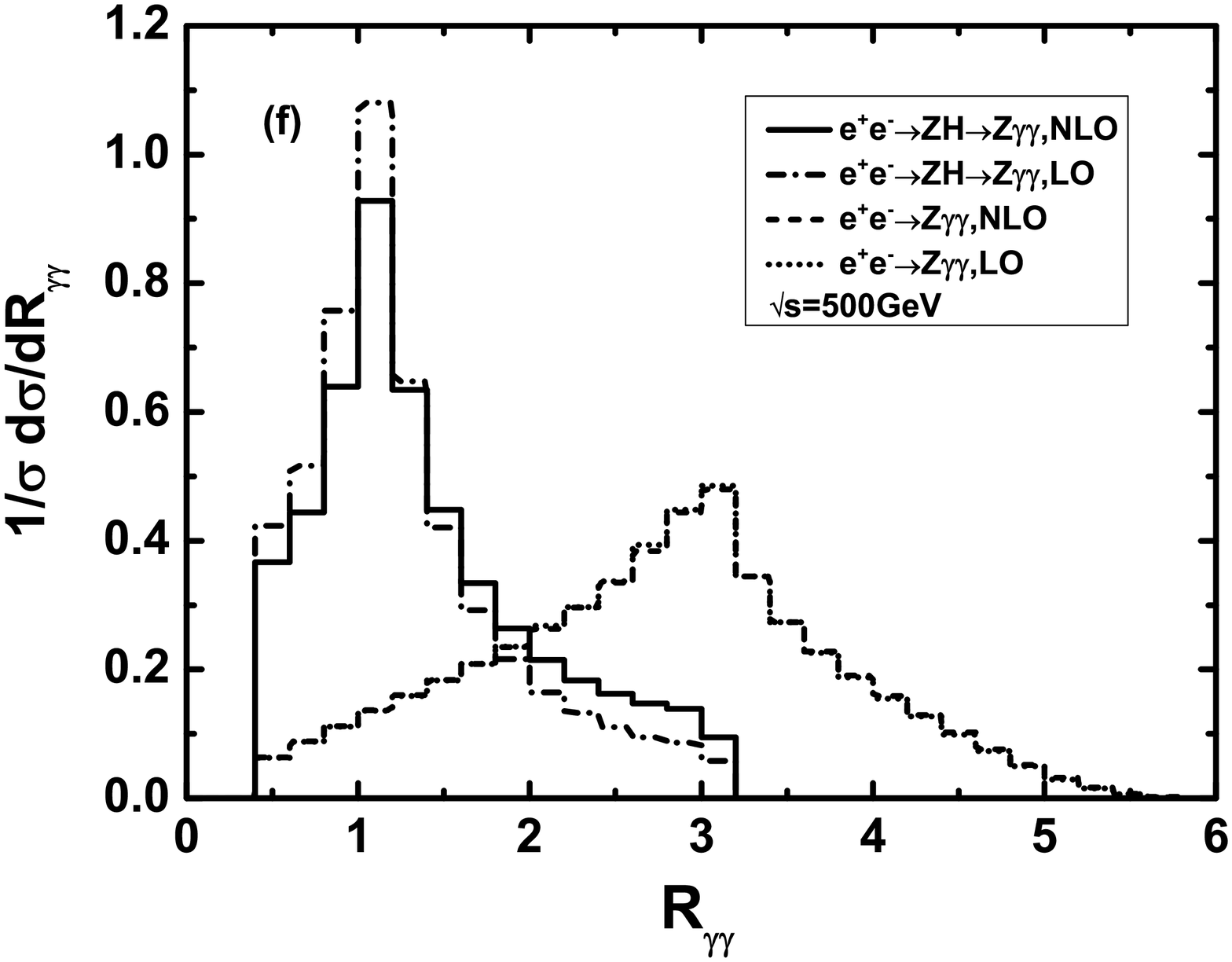}%
\hspace{0in}%
\caption{The normalized kinematic distributions for the signal
process $e^+e^- \to ZH \to Z\gamma\gamma$ and the background \eezrr
process at the $\sqrt{s}=500~GeV$ ILC. All curves in the six figures
are normalized by their total cross sections. (a) Transverse
momentum distributions of final $Z$ boson. (b) Transverse momentum
distributions of the leading photon. (c) Transverse momentum
distributions of the subleading photon. (d) Rapidity distributions
of the leading photon. (e) Rapidity distributions of the subleading
photon. (f) Distributions of the separation $R_{\gamma\gamma}$
between the final leading and subleading photons. }
\label{fig-higgs}
\end{figure}

As we know that one of the most important reactions at the ILC for
Higgs boson precision study is the $e^+e^- \to ZH$ process followed
by $H \to \gamma\gamma$ decay, while this signal process is
accompanied by a serious background process \eezrr. The one-loop
radiative corrections to this signal process $e^+e^- \to ZH$ within
the SM were calculated by A. Denner, {\it et al} \cite{eeZH}. Here
we follow the strategy used in Ref.\cite{eeZH} for the calculation
of the $e^+e^- \to ZH$ process up to the EW NLO within the SM, and we 
adopt the input parameters presented in our work (see Section III.1)
to calculate the LO and NLO EW corrected results for the $e^+e^- \to
ZH \to Z\gamma\gamma$ signal process. The decay width of the SM
Higgs is obtained by using the program HDECAY \cite{hdecay}. Since
the kinematics of the signal events is distinctively different from
that of background events. This difference can be used to suppress
the background and enhance the ratio of signal to background (S/B).
Taking advantage of the kinematic difference, we expect that we can
impose the optimal cuts to extract the signal $e^+e^-\to ZH\to
Z\gamma\gamma$ from the SM background \eezrr efficiently. To
illustrate the distribution differences between the signal and the
background, we present the normalized LO and NLO EW corrected
distributions of various kinematic observables of the final particles
for the signal process $e^+e^-\to ZH\to Z\gamma\gamma$ and the
background process \eezrr in Figs.\ref{fig-higgs}(a-f). All the
results are presented in conditions of $\sqrt s = 500~GeV$,
$p^{\gamma}_T \ge 15~GeV$, $|y_{\gamma}|\le 2.5$,
$R_{\gamma\gamma}\ge 0.4$, and the 'inclusive' event selection
scheme. We show the $p_T$ distributions of the final $Z$-boson,
leading photon and  subleading photon in Figs.\ref{fig-higgs}(a-c),
separately. We can see that, compared to the background, the typical
feature of the signal is that the final state particles are
distributed in the large transverse momentum regions, especially for
the final $Z$-boson and the leading photon. The normalized rapidity
distributions of the final leading and subleading photons are
demonstrated in Fig.\ref{fig-higgs}(d) and Fig.\ref{fig-higgs}(e),
respectively. It shows that the leading and subleading photons from
Higgs boson decay, mainly appear in the central rapidity region,
while the corresponding distributions for the background process are
rather flat. We can see also from Fig.\ref{fig-higgs}(f) that the
leading and subleading photons produced in the background are more
dramatically separated than in the signal process $e^+e^- \to ZH \to
Z\gamma\gamma$. From all of the Figs.\ref{fig-higgs}(a-f),
we can conclude that if we take some proper cuts on kinematic
variables of the the final $Z$-boson and photons, the background
from the \eezrr process can be significantly suppressed.
%%%%%%%%%%%%%%%%%%figures%%%%%%%%%%%%%%%%%%%%%%

\vskip 5mm
\section{Summary}
\par
The $e^+e^- \to Z \gamma\gamma~$ process is very important for
understanding the nature of the Higgs boson and searching for new
physics beyond the SM. In this work we report on our calculation of
the full NLO EW contributions to the $e^+e^- \to Z \gamma\gamma~$
process in the SM, and we analyze the EW quantum effects on the total
cross section and the kinematic distributions of the final particles. We
study the dependence of the $Z\gamma\gamma$ production rate on the event
selection scheme and provide distributions of some important
observables. We find that the full NLO EW corrections can enhance
the LO total cross sections quantitatively from $2.32\%$ to $9.61\%$
when colliding energy goes up from $250~GeV$ to $1~TeV$, and the
size of the NLO correction exhibits a strong dependence on the
observable and on phase space. We conclude that in studying the
signal process $e^+e^- \to ZH \to Z \gamma\gamma~$, the background
events of $e^+e^- \to Z \gamma\gamma~$ process can be suppressed
significantly if we take appropriate kinematic cuts on the final
products.

\vskip 5mm
\par
\noindent{\large\bf Acknowledgments:} This work was supported in
part by the National Natural Science Foundation of China
(Grants. No.11275190, No.11375008, No.11375171), and the Fundamental
Research Funds for the Central Universities (Grant. No.WK2030040044).

\end{document}